\documentclass[prl,amsmath,amssymb,twocolumn,floatfix,superscriptaddress]{revtex4}
\usepackage{graphicx}
\usepackage{times}
\usepackage{xcolor}

\begin{document}
\title{Seismic Invisibility: Elastic wave cloaking via symmetrized transformation media}

\author{Sophia R. Sklan}
\affiliation{Department of Mechanical Engineering, University of Colorado Boulder,
Colorado 80309 USA}
\author{Ronald Y.S. Pak}
\affiliation{Department of Civil Engineering, University of Colorado Boulder,
Colorado 80309 USA}
\author{Baowen Li}
\affiliation{Department of Mechanical Engineering, University of Colorado Boulder,
Colorado 80309 USA}

\begin{abstract}
Transformation media theory, which steers waves in solids via an effective geometry induced by a refractive material (Fermat's principle of least action), provides a means of controlling vibrations and elastic waves beyond the traditional dissipative structures regime.
In particular, it could be used to create an elastic wave cloak, shielding an interior region against elastic waves while simultaneously preventing scattering in the outside domain.
\textcolor{black}{However, as a true elastic wave cloak would generally require materials with stiffness tensors lacking the minor symmetry (implying asymmetric stress), the utility of such an elastic wave cloak has thus far been limited by the challenge of fabricating these materials}.
Here we develop a means of overcoming this limitation via the development of a symmetrized elastic cloak, sacrificing some of the performance of the perfect cloak for the sake of restoring the minor symmetry.
We test the performance of the symmetrized elastic cloak for shielding a tunnel against seismic waves, showing that it can be used to reduce the average displacement within the tunnel by an order of magnitude (and reduce energy by two orders of magnitude) for waves above a critical frequency of the cloak.
This critical frequency, which corresponds to the generation of surface waves at the cloak-interior interface, can be used to develop a simple heuristic model of the symmetrized elastic cloak's performance for a generic problem.
\end{abstract}


\maketitle

The need to protect objects against unwanted mechanical vibration and wave incidence is a long-standing subject in engineering \cite{PVC,SD}.
The problem arises in multiple scenarios including blocking sound/elastic waves and eliminating unwanted mechanical resonances \cite{PVC,SD} (vibration isolation) or guarding against nonlinear mechanical shock waves and preventing the collapse of structures under incident seismic waves \cite{Hist,Towhata} (earthquake engineering).
Techniques to accomplish these goals have included strengthening structures, modifying structural resonances through additive engineering or anti-resonance, developing flexible structures that can withstand large deformations, or including dissipative elements \cite{PVC,SD,Hist,Towhata}.
Recently, techniques from phononics and acoustic wave engineering have been adapted to vibration isolation/earthquake engineering (VIEE), most prominently by the inclusion of phononic/sonic crystals as a means of shielding against seismic surface waves \cite{PVC,Towhata,SPC1,SPC2,SPC3,SPC4,SPC5,SPC6,SPC7} (which contain a much smaller portion of energy from an earthquake than bulk waves, particularly shear waves \cite{EE}).
However, one limitation of all these techniques is that they can only seek to mitigate elastic waves, reducing the local amplitude of the earthquake or allowing a structure to withstand an unmodified vibration.
In addition, many of these techniques have focused upon narrow frequency bands (e.g. modifying resonances), particularly for low frequency applications.
While these are primary concerns, minimization of broadband or higher frequency transmissions to a structure  is of increasing engineering interest nowadays as the reliance on electronic-computer control of critical equipment grows.   


An alternative framework to VIEE would be cloaking.
Cloaks modify the environment around a region such that waves are refracted around a central domain \cite{TO1,TO2,RC1,CC1,CC2,CC3}.
Since the energy is redirected, not dissipated, it could in principle be used to control vibrations of arbitrary amplitude or frequency (in practice, engineering limitations prevent perfect performance).
Developing effective, realizable cloaks is an active field of research in optics and acoustics, but application to elastic waves and VIEE has remained in its infancy \cite{EC1,EC2,EC3,EC4,EC5,EC6,EC7}.
The central challenge limiting the utilization of cloaking for these applications is that transformation media theory, the mathematical framework underlying the operation of the cloak, is not feasible for a generic elastic wave.
As was shown by Milton et al., a cloaking transformation for an elastic wave would break the stiffness tensor's minor symmetry ($c_{ijkl}=c_{jikl}=c_{ijlk}$) \cite{Milton}.
Since this symmetry exists for all \textcolor{black}{commonplace solid} materials, this has limited elastic wave cloaks to special cases where the loss of minor symmetry is irrelevant (e.g. planar structures,  partial coordinate transformations) or \textcolor{black}{structures with significant fabrication challenges (e.g. Cosserat materials and auxetic/pentamode materials, hyperelastic metamaterials, Willis materials) \cite{AEC1,AEC2,AEC3,AEC4,AEC5,AEC6,AEC7,AEC8,AEC9,AEC10,AEC11}.
Hyperelastic metamaterials \cite{HEM} deserve special mention as a system where the nonlinear pre-stress of the system does allow for the more ready construction of materials lacking minor elastic symmetry.
However, fabrication challenges and the need for strong nonlinearity still limit the widespread application of the technique.
Similarly, Willis materials \cite{Milton} provide an explicit solution where the loss of minor symmetry is preserved, but at the cost of introducing highly dispersive materials which distort the waves and thus limit the utility of broadband cloaking.
Moreover, the incorporation of Willis materials at a theoretical level has not noticeable reduced the fabrication challenges of constructing an elastic wave cloak.}

In this work we present a framework for an approximate elastic cloak, which preserves minor symmetries even in the most general case.
This symmetrized elastic cloak (SEC) has the advantage of being, in principle, \textcolor{black}{more readily} realizable, but comes at the cost of no longer being a perfect cloak.
However, a cloak generally performs two tasks: preventing scattering of an incoming wave as it propagates around and through the cloak (stealth, the primary concern in optical or acoustic cloaking) and blocking waves from penetrating a central region (shielding, the primary concern in VIEE).
The performance of an approximate cloak for both these tasks is an open question, but only the latter performance metric is important for VIEE applications.
As such, we characterize the performance of the cloak in the simplest physically realistic scenario $-$ shielding a tunnel or a round shell buried in a soil-type medium from seismic waves.
Through this analysis, we derive the limitations of the SEC and present a simple holistic model that captures the essential performance characteristics.

To begin, the equations of motion in Cartesian coordinates for elastic waves in a solid are
\begin{eqnarray}
\rho\partial_{tt}u_i&=&\partial_j\sigma_{ji} \\
\sigma_{ij}&=&c_{ijkl}\epsilon_{kl}\\
\epsilon_{ij}&=&\frac{1}{2}(\partial_ju_i+\partial_iu_j),
\end{eqnarray}
(where $\rho$ is density, $u$ is the displacement vector, $\epsilon$ is the infinitesimal strain tensor, $\sigma$ is the Cauchy stress tensor, and $c$ is the 4th order stiffness tensor).
In cylindrical coordinates they become
\begin{eqnarray}\label{eq:EOM}
\rho\partial_{tt}u_r&=&\frac{1}{r}\partial_rr\sigma_{rr}+\frac{1}{r}\left(\partial_\theta\sigma_{\theta r}-\sigma_{\theta\theta}\right)+\partial_z\sigma_{zr}\nonumber\\
\rho\partial_{tt}u_\theta&=&\frac{1}{r}\partial_rr\sigma_{r\theta}+\frac{1}{r}\left(\partial_\theta\sigma_{\theta\theta}+\sigma_{\theta r}\right)+\partial_{z}\sigma_{z\theta}\\
\rho\partial_{tt}u_z&=&\frac{1}{r}\partial_rr\sigma_{rz}+\frac{1}{r}\partial_\theta\sigma_{\theta z}+\partial_z\sigma_{zz}, \nonumber
\end{eqnarray}
while $\sigma_{ij}=c_{ijkl}\epsilon_{ij}$ is unchanged except for a relabeling of coordinates, and 
\begin{equation}\label{eq:eps}
\overset{=}{\epsilon}=\left[\begin{array}{ccc}
\partial_ru_r\hat{e}_r\hat{e}_r & \frac{1}{r}(\partial_\theta u_r-u_{\theta})\hat{e}_r\hat{e}_\theta & \partial_zu_r\hat{e}_r\hat{e}_z\\
\partial_ru_\theta\hat{e}_\theta\hat{e}_r & \frac{1}{r}(\partial_\theta u_\theta+u_r)\hat{e}_\theta\hat{e}_\theta & \partial_zu_\theta\hat{e}_\theta\hat{e}_z\\
\partial_ru_z\hat{e}_z\hat{e}_r & \frac{1}{r}\partial_\theta u_z\hat{e}_z\hat{e}_\theta & \partial_zu_z\hat{e}_z\hat{e}_z
\end{array}\right]
\end{equation}
where we've dropped the explicit symmetry of $\epsilon$ and $\sigma$ by requiring it be preserved in $c$.
In general, $c$ must possess both the major symmetry
\begin{equation}
c_{ijkl}=c_{klij}
\end{equation}
(which comes from the symmetry of mixed partials) and the minor symmetry
\begin{equation}
c_{ijkl}=c_{jikl}=c_{ijlk}=c_{jilk}
\end{equation}
(which comes from the physical symmetry of stress and strain).
The standard cylindrical cloaking transformation \cite{TO1} is the dilation
\begin{equation}
r^\prime=\frac{b-a}{b}r+a
\end{equation}
where $r^\prime$ is the transformed radial coordinate, $a$ is the inner radius of the cloak, and $b$ is the outer radius of the cloak.
Under the cloaking transformation, we seek a set of $c^\prime$ and $\rho^\prime$ such that the equation of motion in the standard frame (i.e. equation (\ref{eq:EOM})) with these materials is the same as the equation of motion in the transformed frame (equation (\ref{eq:EOM}) with all factors of $r$ replaced by $r^\prime$ but $\vec{u}^\prime=\vec{u}$ and $\sigma^\prime\ne\sigma$) with some trivial background material.
Breaking the minor symmetry, this can be accomplished exactly using a density and stiffness tensor
\begin{equation}\label{eq:rhoTr}
\rho^\prime=\rho_0\frac{r-a}{r}\left(\frac{b}{b-a}\right)^2
\end{equation}
\begin{widetext}
\begin{gather}
c_{ijkl}^{\prime}=\left(\begin{array}{ccc}
\frac{r-a}{r}\\
 & 1\\
 &  & \frac{b}{b-a}\frac{r-a}{r}\end{array}\right)\times\nonumber\\ \label{eq:CTr}
\left(\begin{array}{ccc}
\left[\begin{array}{ccc}
c_{rrrr} & c_{rrr\theta} & c_{rrrz}\\
\frac{r}{r-a}c_{rr\theta r} & \frac{r}{r-a}c_{rr\theta\theta} & \frac{r}{r-a}c_{rr\theta z}\\
\frac{b}{b-a}c_{rrzr} & \frac{b}{b-a}c_{rrz\theta} & \frac{b}{b-a}c_{rrzz}\end{array}\right] & \left[\begin{array}{ccc}
c_{r\theta rr} & c_{r\theta r\theta} & c_{r\theta rz}\\
\frac{r}{r-a}c_{r\theta\theta r} & \frac{r}{r-a}c_{r\theta\theta\theta} & \frac{r}{r-a}c_{r\theta\theta z}\\
\frac{b}{b-a}c_{r\theta zr} & \frac{b}{b-a}c_{r\theta z\theta} & \frac{b}{b-a}c_{r\theta zz}\end{array}\right] & \left[\begin{array}{ccc}
c_{rzrr} & c_{rzr\theta} & c_{rzrz}\\
\frac{r}{r-a}c_{rz\theta r} & \frac{r}{r-a}c_{rz\theta\theta} & \frac{r}{r-a}c_{rz\theta z}\\
\frac{b}{b-a}c_{rzzr} & \frac{b}{b-a}c_{rzz\theta} & \frac{b}{b-a}c_{rzzz}\end{array}\right]\\
\left[\begin{array}{ccc}
c_{\theta rrr} & c_{\theta rr\theta} & c_{\theta rrz}\\
\frac{r}{r-a}c_{\theta r\theta r} & \frac{r}{r-a}c_{\theta r\theta\theta} & \frac{r}{r-a}c_{\theta r\theta z}\\
\frac{b}{b-a}c_{\theta rzr} & \frac{b}{b-a}c_{\theta rz\theta} & \frac{b}{b-a}c_{\theta rzz}\end{array}\right] & \left[\begin{array}{ccc}
c_{\theta\theta rr} & c_{\theta\theta r\theta} & c_{\theta\theta rz}\\
\frac{r}{r-a}c_{\theta\theta\theta r} & \frac{r}{r-a}c_{\theta\theta\theta\theta} & \frac{r}{r-a}c_{\theta\theta\theta z}\\
\frac{b}{b-a}c_{\theta\theta zr} & \frac{b}{b-a}c_{\theta\theta z\theta} & \frac{b}{b-a}c_{\theta\theta zz}\end{array}\right] & \left[\begin{array}{ccc}
c_{\theta zrr} & c_{\theta zr\theta} & c_{\theta zrz}\\
\frac{r}{r-a}c_{\theta z\theta r} & \frac{r}{r-a}c_{\theta z\theta\theta} & \frac{r}{r-a}c_{\theta z\theta z}\\
\frac{b}{b-a}c_{\theta zzr} & \frac{b}{b-a}c_{\theta zz\theta} & \frac{b}{b-a}c_{\theta zzz}\end{array}\right]\\
\left[\begin{array}{ccc}
c_{zrrr} & c_{zrr\theta} & c_{zrrz}\\
\frac{r}{r-a}c_{zr\theta r} & \frac{r}{r-a}c_{zr\theta\theta} & \frac{r}{r-a}c_{zr\theta z}\\
\frac{b}{b-a}c_{zrzr} & \frac{b}{b-a}c_{zrz\theta} & \frac{b}{b-a}c_{zrzz}\end{array}\right] & \left[\begin{array}{ccc}
c_{z\theta rr} & c_{z\theta r\theta} & c_{z\theta rz}\\
\frac{r}{r-a}c_{z\theta\theta r} & \frac{r}{r-a}c_{z\theta\theta\theta} & \frac{r}{r-a}c_{z\theta\theta z}\\
\frac{b}{b-a}c_{z\theta zr} & \frac{b}{b-a}c_{z\theta z\theta} & \frac{b}{b-a}c_{z\theta zz}\end{array}\right] & \left[\begin{array}{ccc}
c_{zzrr} & c_{zzr\theta} & c_{zzrz}\\
\frac{r}{r-a}c_{zz\theta r} & \frac{r}{r-a}c_{zz\theta\theta} & \frac{r}{r-a}c_{zz\theta z}\\
\frac{b}{b-a}c_{zzzr} & \frac{b}{b-a}c_{zzz\theta} & \frac{b}{b-a}c_{zzzz}\end{array}\right]\end{array}\right).
\end{gather}
\end{widetext}
Note that this preserves major symmetry of $c$ and that the transformation of $c_{ijkl}$ is effectively independent of two indices (either $i$ or $j$ and either $k$ or $l$).

Because equation (\ref{eq:CTr}) has lost minor symmetry, however, it is no longer \textcolor{black}{easily} realizable.
To restore this symmetry, we impose a symmetrization function
\begin{equation}
c^S_{ijkl}=S(c_{ijkl},c_{jikl},c_{ijlk},c_{ijilk})
\end{equation}
where $S$ is an arbitrary function that preserves minor symmetry.
\textcolor{black}{(Note that a symmetrization technique was also applied in Ref. \cite{Sct}, but as the focus there was on the scattering field and the focus here is on shielding, their results and analysis differ from ours.}
\textcolor{black}{In particular, when Ref. \cite{Sct} considers cloaking, they measure the perfomance of the cloak with respect to the external scattered field.
That is, their efficiency measure is the normalized r.m.s. deviation of the external field with respect to a homogeneous background.
Any effect on the fields in the cloaked region is explicitly excluded from their analysis, following Ref. \cite{Eff}.
This differs from our work, which considers the performance of the cloak in shielding an internal structure from an external source.
As such, we focus on the deviation of the internal field with respect to an uncloaked object and ignore any effects on the external, scattered field.
Thus, the focus of our work is complementary to Ref. \cite{Sct} and the conclusions we draw on the efficacy of symmetrized cloaks to shield vibrations does not conflict with the conclusions they draw on the efficacy of symmetrized cloaks to prevent scattering (i.e. that an anisotropic density tensor is a necessary condition to create a cloak that effectively reduces scattering while incorporating a symmetrized elasticity tensor).)
}
For $c^S$'s simplicity, we select the geometric mean:
\begin{equation}
S_{GM}(x,y,z,w)=(xyzw)^{1/4}
\end{equation}
 for our SEC.
(In principle some other symmetrization function could improve the performance of the cloak under some criteria, although it is intuitive that some average function is optimal.)
With this symmetry imposed, our stiffness tensor is reduced to the 21 components (in Voigt notation, $rr=1,\theta\theta=2,zz=3,\theta z=4,\theta r=5,r\theta=6$)
\begin{widetext}\begin{equation}
c_{IJ}^{S}=\left[\begin{array}{cccccc}
\frac{r-a}{r}c_{11} & c_{12} & \frac{b}{b-a}\frac{r-a}{r}c_{13} & (\frac{b}{b-a}\frac{r-a}{r})^{1/2}c_{14} & (\frac{b}{b-a})^{1/2}\frac{r-a}{r}c_{15} & (\frac{r-a}{r})^{1/2}c_{16}\\
 & \frac{r}{r-a}c_{22} & \frac{b}{b-a}c_{23} & (\frac{b}{b-a}\frac{r}{r-a})^{1/2}c_{24} & (\frac{b}{b-a})^{1/2}c_{25} & (\frac{r}{r-a})^{1/2}c_{26}\\
 &  & (\frac{b}{b-a})^{2}\frac{r-a}{r}c_{33} & (\frac{b}{b-a})^{3/2}(\frac{r-a}{r})^{1/2}c_{34} & (\frac{b}{b-a})^{3/2}\frac{r-a}{r}c_{35} & \frac{b}{b-a}(\frac{r-a}{r})^{1/2}c_{36}\\
 &  &  & \frac{b}{b-a}c_{44} & (\frac{b}{b-a})^{1/2}(\frac{r-a}{r})^{1/4}c_{45} & (\frac{b}{b-a})^{1/4}c_{46}\\
 &  &  &  & \frac{b}{b-a}\frac{r-a}{r}c_{55} & (\frac{b}{b-a}\frac{r-a}{r})^{1/2}c_{56}\\
 & S & Y & M &  & c_{66}\end{array}\right]
\end{equation}\end{widetext}
(the lower half of the tensor omitted by symmetry).
For a cloak embedded within an isotropic background, we can impose a further simplification to 
\begin{widetext}\begin{equation}\label{eq:SEC}
c^{S}(\lambda,\mu)=\left[\begin{array}{cccccc}
\frac{r-a}{r}(\lambda+2\mu) & \lambda & \frac{b}{b-a}\frac{r-a}{r}\lambda\\
\lambda & \frac{r}{r-a}(\lambda+2\mu) & \frac{b}{b-a}\lambda\\
\frac{b}{b-a}\frac{r-a}{r}\lambda & \frac{b}{b-a}\lambda & (\frac{b}{b-a})^{2}\frac{r-a}{r}(\lambda+2\mu)\\
 &  &  & \frac{b}{b-a}\mu\\
 &  &  &  & \frac{b}{b-a}\frac{r-a}{r}\mu\\
 &  &  &  &  & \mu\end{array}\right].
\end{equation}\end{widetext}
Notably, this implies that the stiffness tensor for a cloak in an isotropic material should possess orthotropic symmetry.

\begin{figure}\begin{center}
\includegraphics[scale=0.6]{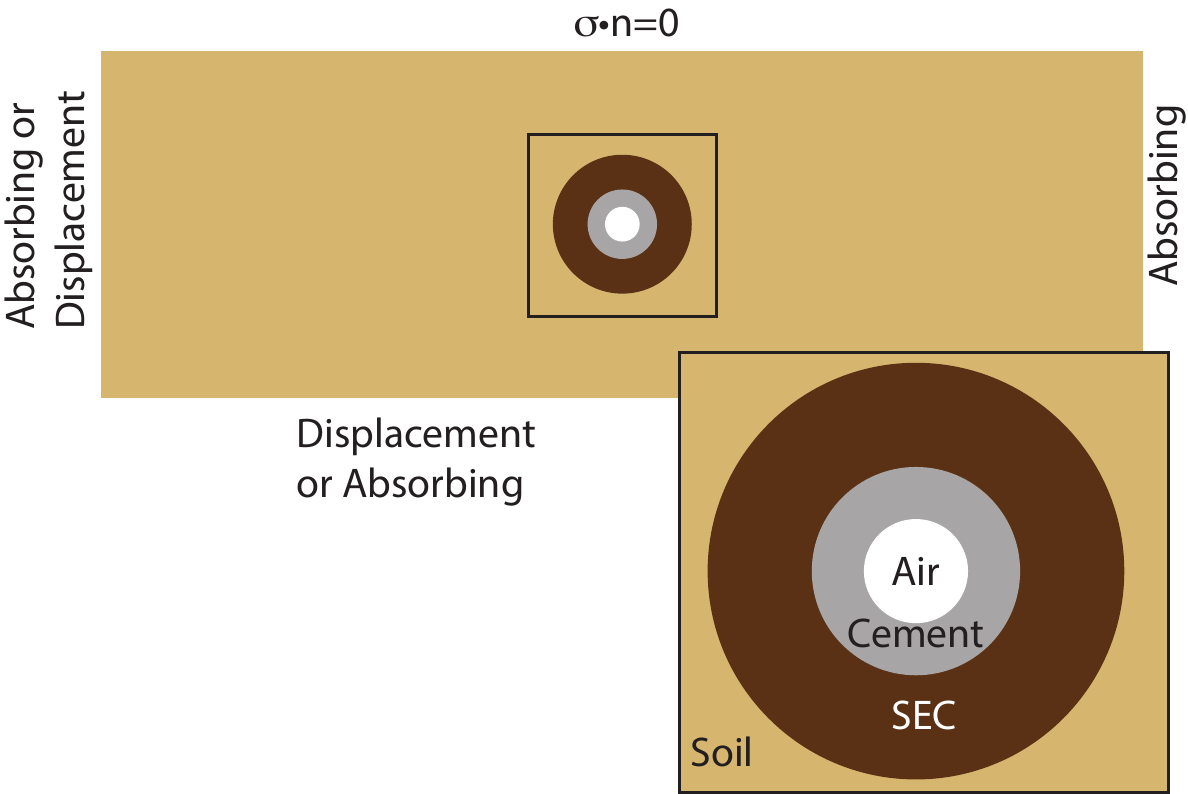}
\caption{\label{fig:model}Schematic model of SEC tunnel shield simulation setup. The cloak and tunnel are placed at the center of simulation domain, with boundary conditions labeled at each end. (Inset) Material structure of the simulation: Soil (light brown) surrounds the SEC (dark brown ring). Inside the SEC is a concrete (grey ring) wall filled with air (white circle).}
\end{center}\end{figure}

While this mathematical framework for the SEC is consistent $-$ it is a set of material properties which mimic the perfect cloak in some respects but maintain \textcolor{black}{the symmetries of common materials} $-$ its utility is a separate matter.
After all, when our material properties deviate from the perfect cloak its performance will degrade.
To test the SEC's performance, it is necessary to impose a test case and a metric.
Since our goal for the SEC is VIEE applications, we shall test it by specializing to seismic waves and considering the highest symmetry case, namely protecting a hollow core concrete tunnel (concrete: $\rho=2300\mathrm{kg/m^3},\mu=9.34\mathrm{GPa},\lambda=9.12\mathrm{GPa}$, air: $\rho=1.225\mathrm{kg/m^3},\mu=0\mathrm{Pa},\lambda=142\mathrm{kPa}$) buried in uniform soil ($\rho=2203\mathrm{kg/m^3},\mu=30\mathrm{MPa},\lambda=120\mathrm{MPa}$) with a flat surface.
For waves \textcolor{black}{incident} orthogonal to the length of the tunnel, then, we can approximate the system as two-dimensional (see Fig. \ref{fig:model}).
We consider a tunnel of radius $a=$10m with a hollow center of 5m radius (in practice, this concrete shell is likely thicker than a real tunnel, but was utilized for numerical stability) and an inhomogeneous cloak obeying equations (\ref{eq:rhoTr}), (\ref{eq:CTr}) with radius $b=$20m (this implies an SEC thickness of 10m, which is likely much thicker than required in practice $-$ the thickness of the cloak being chiefly constrained by the feature size of the metamaterials used in its fabrication).
This tunnel was placed in the center of a rectangular domain (100m$\times$300m) of soil.
For boundary conditions, the top surface was left traction-free ($\sigma\cdot\hat{n}=0$), with the sides left as impedance matched absorbing layers.
\textcolor{black}{(Note that a true impedance matching requires breaking the minor symmetry \cite{Sct}.
Instead, we only use the pre-defined ``low reflecting boundary'' setting of COMSOL \cite{COMSOL}, which imposes a boundary that only approximately impedance matches every polarization and will introduce some amount of reflection.
The simulation domain's boundaries were set to minimize the impact of these reflections on the performance of the SEC.)}
From the base, a vibration of $u_i=u_{i0}\sin(2\pi\omega t)\hat{e}_i$ was imposed, with $u_{i0}=$1cm and $i=x,y$ for S or P waves respectively.
Frequencies of the imposed waves were allowed to vary between 1 and 10 Hz, as those are the most relevant frequencies for typical earthquakes \cite{HFW}.
In addition, a second set of boundary conditions are considered where waves approach parallel to the surface, i.e. the displacement is imposed upon one of the sides and the base is left impedance matched.
\textcolor{black}{This second set of boundary conditions was used to account for any effects induced by the anisotropy of the boundary conditions; the operation of the cloak in infinite homogeneous space is expected to retain isotropy.}
To measure the efficiency of the cloak, we employ two metrics: reduction in average energy and reduction in average displacement.
These are characterized as efficiencies,
\begin{equation}
\eta(E)=1-\langle E\rangle_C/\langle E\rangle_U
\end{equation}
and
\begin{equation}
\eta({u_i})=1-\langle |u_i|\rangle_C/\langle |u_i|\rangle_U
\end{equation}
where brackets denote averaging over the concrete shell and averaging over time, $E$ is total energy, $U$ refers to results with no cloak, and $C$ refers to results with the cloak.
Given that we consider two directions of incident waves (other directions can be decomposed into a linear superposition of these orthogonal cases), we define $u_\Vert$ as the response with polarization parallel to the driving force and $u_\perp$ as the transverse response (in general $u_\Vert\gg u_\perp$, $\eta(u_\perp),\eta(v_i),\eta(a_i)$ in supplement).
Note that $\eta\le1$ for a passive cloak, with $\eta=1$ being perfect cloaking (100$\%$ efficiency), $\eta=0$ being no improvement, and $\eta<0$ being a degradation of performance due to the cloak.
With this setup, we use COMSOL \cite{COMSOL} to simulate the tunnel with and without the cloak for varying boundary conditions ($\vec{u}=0$ initial conditions, simulation interval of $180/\omega$).
\textcolor{black}{Simulations used a mesh grid of 9924 domain elements and 388 bounary elements (i.e. an ``extremely fine mesh'' setting), which showed a strong convergence and included a time resolution of $1/20\omega$.
Numerical integration was performed using the pre-built COMSOL finite element differential equation solver using MUMPS at COMSOL's default settings and displacements explicitly constrained to real values.}

\begin{figure}\begin{center}
\includegraphics[scale=0.305]{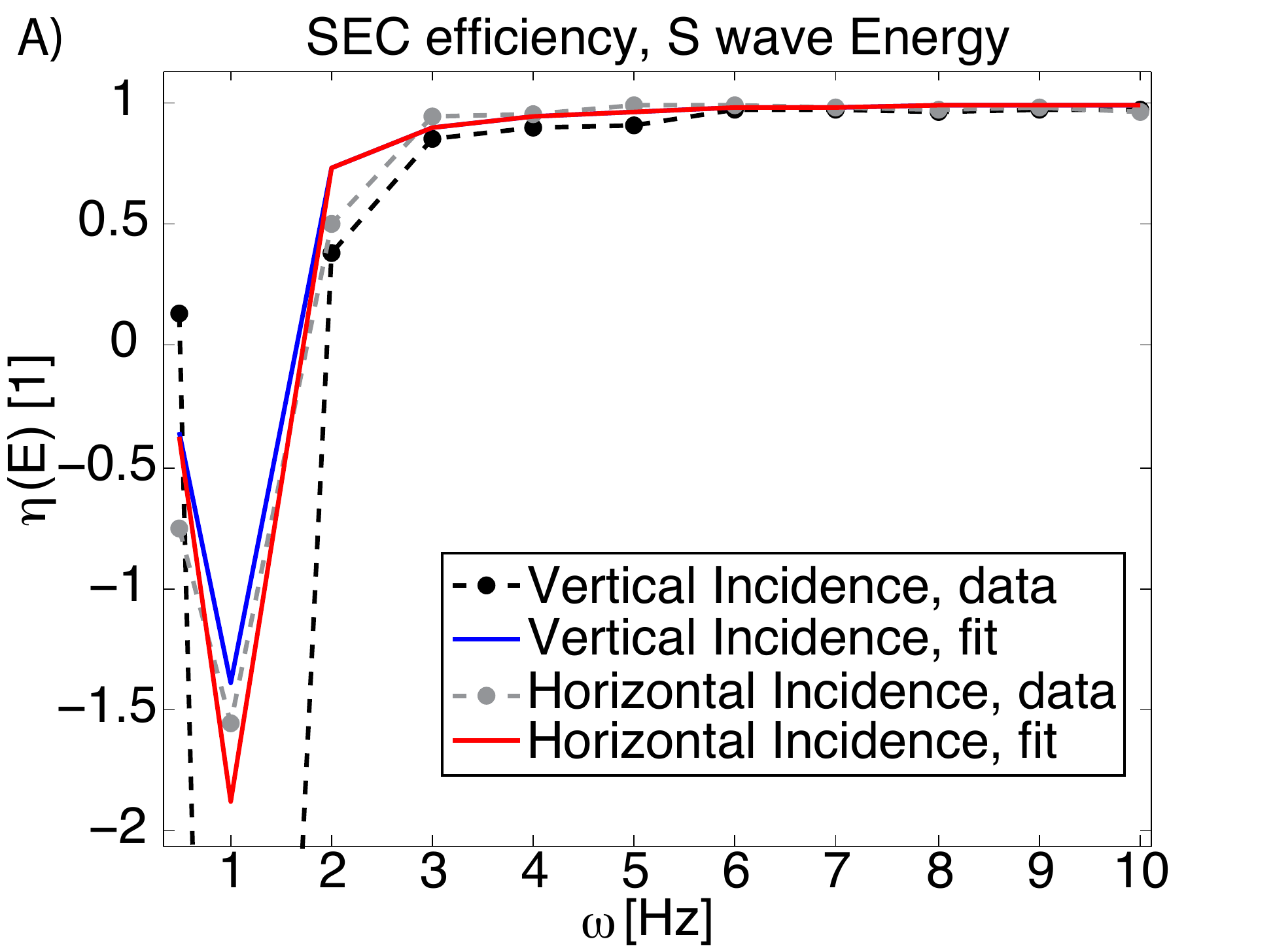}
\includegraphics[scale=0.305]{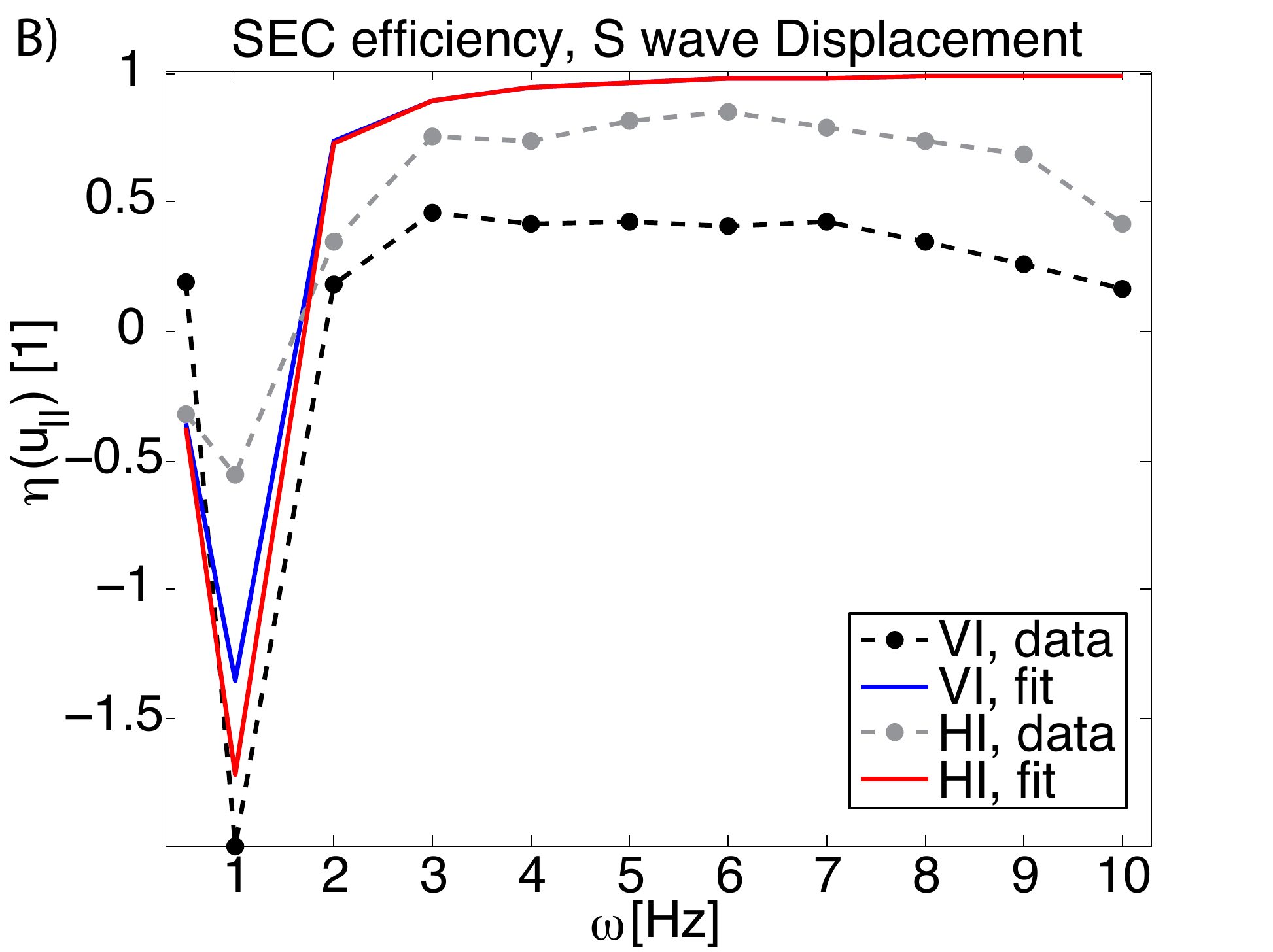}
\includegraphics[scale=0.305]{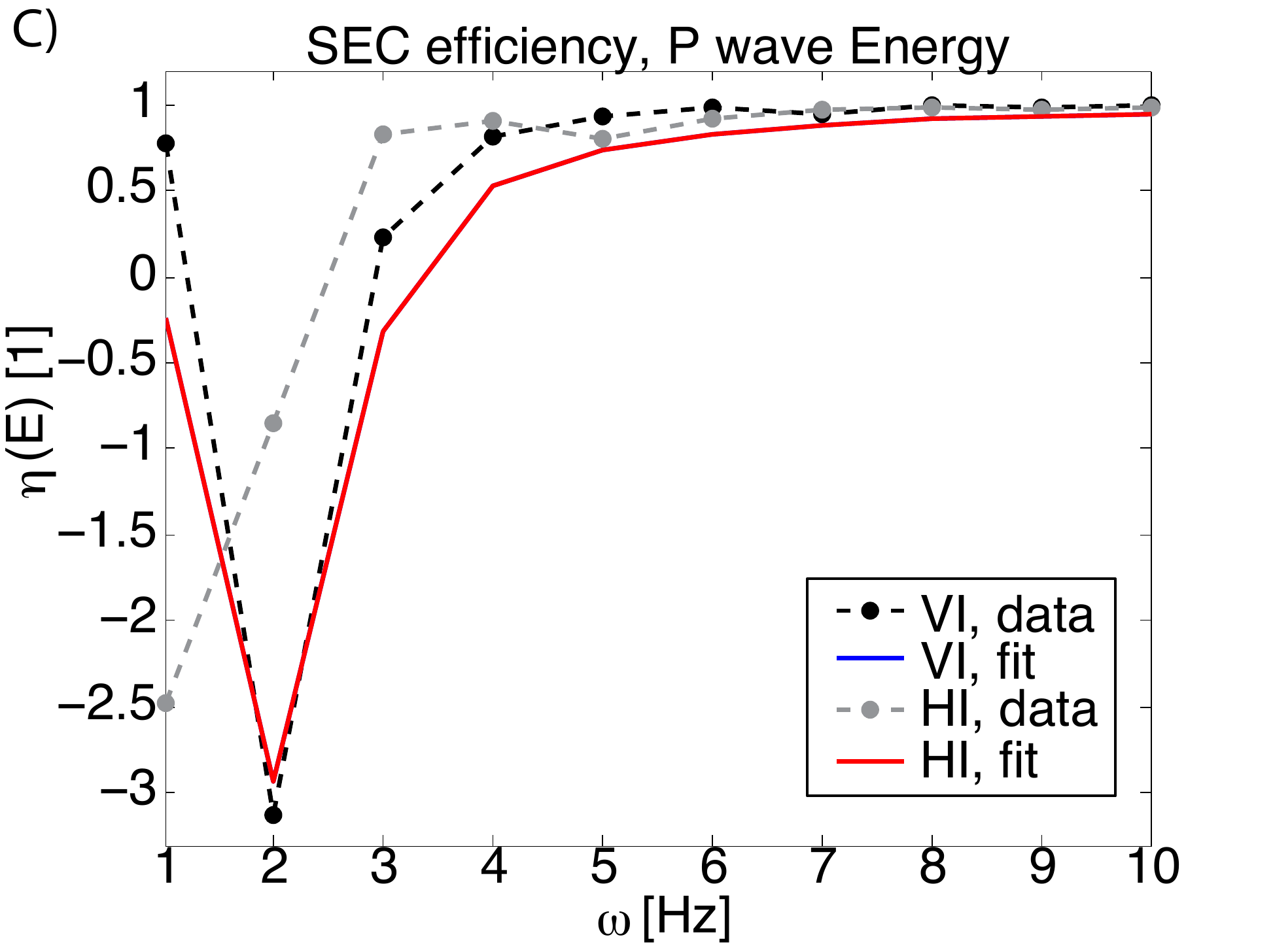}
\includegraphics[scale=0.305]{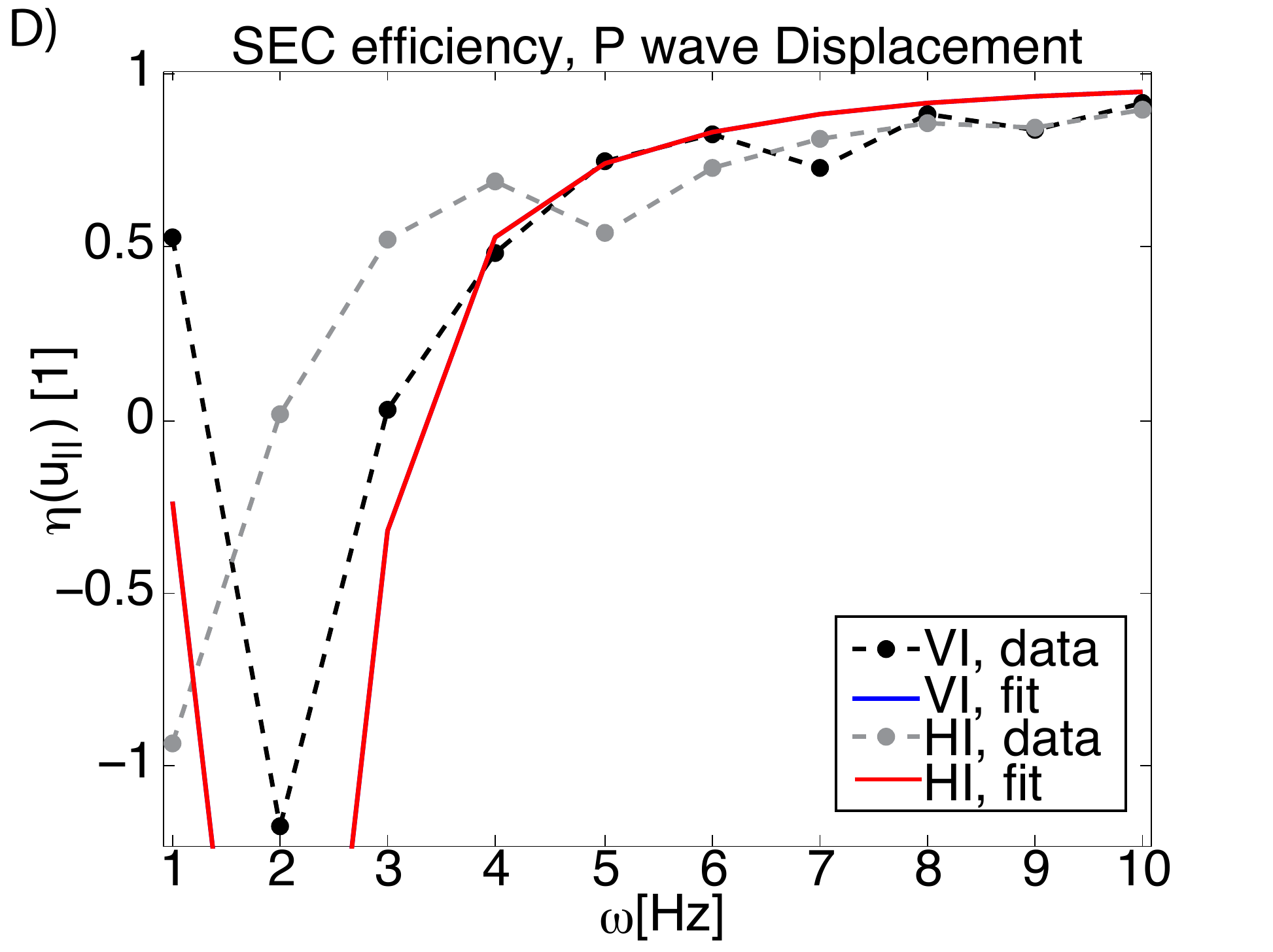}
\caption{\label{fig:eff}Efficiency plots for the SEC as a function of frequency. Black dashed curve is data for vertical incidence (VI, wave from beneath the SEC), grey dashed curve is data for horizontal incidence (HI, wave from the left side of the SEC), while solid blue and red curves are fitted harmonic oscillator models of the SEC for VI and HI respectively. \textcolor{black}{Fits are generated by matching the data generated through direct COMSOL simulations of the dynamics with a simple harmonic oscillator model given in equation (\ref{eq:fit}).} (A) Energy efficiency, S wave. Efficiency at the cutoff frequency of 1Hz (value given in main text) cropped to distinguish higher frequency variation. (B) Longitudinal displacement efficiency, S wave. (C) Energy efficiency, P wave. (D) Longitudinal Displacement efficiency, P wave. \textcolor{black}{Note that fitted curves coincide in (C) and (D) due limited degrees of freedom in the harmonic oscillator model of equation (\ref{eq:fit})}.}
\end{center}\end{figure}

Examining the performance of the SEC under this efficiency metric, we can separate out two different regimes for each polarization.
For S waves (see Figure \ref{fig:eff}A-B), the performance is relatively insensitive to the direction of the incident waves, but shows a clear change in performance above and below approximately 1Hz$\equiv\omega_{c,S}$.
For $\omega\lesssim$1Hz, the SEC clearly reduces the performance of the concrete in blocking incoming waves, reaching its minimum performance at $\omega\approx$1Hz ($\eta(u_{\Vert,S})\gtrsim-2.00,\eta(E_S)\gtrsim-8.11$).

On the other hand, for $\omega\gtrsim$1Hz the SEC shows great improvement in performance; the concrete shell is protected from incoming waves by an order of magnitude for displacement and two orders of magnitude for energy ($\eta(u_{\Vert,S})\lesssim0.849,\eta(E_S)\lesssim0.988$).
P waves (see Figure \ref{fig:eff}C-D) are similar to S waves but the critical frequency separating these regimes is shifted to 2Hz$\equiv\omega_{c,P}$ ($\eta(u_{\Vert,P})\lesssim0.913,\eta(E_P)\lesssim0.993$) due to the difference in speeds $v_P/v_S=\sqrt{6}\approx$2.4 (where $v_i$ is the velocity, $\sqrt{\mu/\rho}$ for S waves and $\sqrt{(\lambda+2\mu)/\rho}$ for P waves).
Note that, for $\omega$ under the critical frequency, some of the simulations demonstrated small oscillations between positive and negative efficiency.
However, as simulations far below the critical frequency possess wavelengths on the order of kilometers or longer, and the cloak is on the order of meters, we expect that these effects are highly sensitive to boundary conditions and may be numerical artifacts.

\begin{figure}\begin{center}
\includegraphics[scale=0.425]{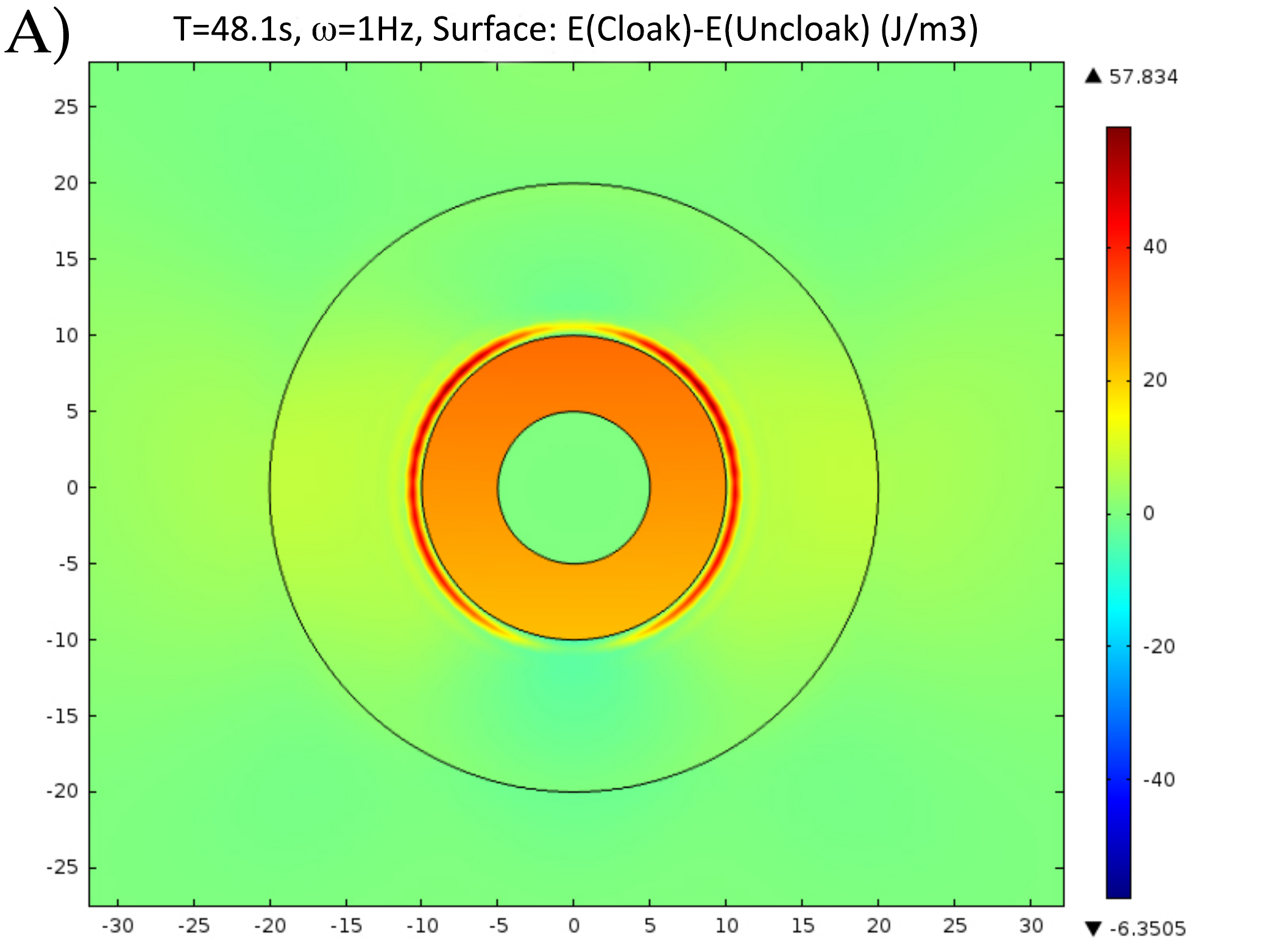}
\includegraphics[scale=0.425]{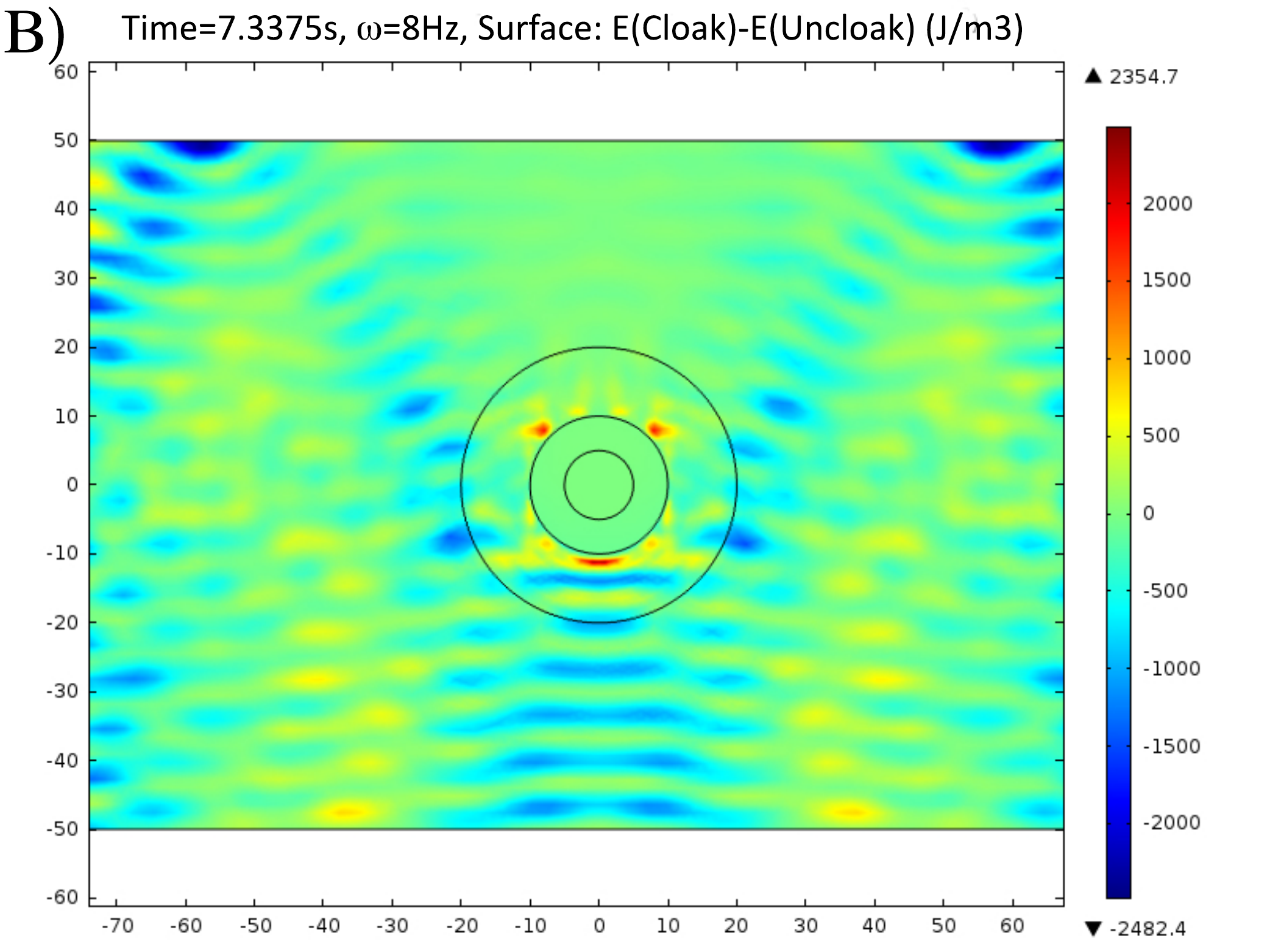}
\caption{\label{fig:sim}Time domain simulations of SEC for VI S waves. Surface plot is energy difference $E_C-E_U$, i.e. positive values (reds) denote lower efficiency and negative values (blues) denote higher efficiency. \textcolor{black}{Field generated by post-processing the displacement field $u(x,t)$ from dynamical simulations in COMSOL. Simulations were paired, with $E_C$ and $E_U$ generated from different instantiations using identical boundary and initial conditions but with the SEC present or absent respectively. This allowed for the generation of the scattered energy field without any explicit scattering field methodology.} (A) Simulation at resonance, $\omega=$1Hz. Note buildup of energy within the concrete tunnel (inner annulus) and the surface wave surrounding it. (B) Simulation above resonance, $\omega=$8Hz. Note the surface wave is localized and does not surround the tunnel. View is zoomed out compared to (A) to display scattering field variation.}
\end{center}\end{figure}

To explain the difference in performance above and below these critical frequencies, we examine the energy difference between the cloaked and uncloaked cases as a function of time for different incident frequencies (see Figure \ref{fig:sim}).
Below the critical frequency (Figure \ref{fig:sim}A), we see a very clear buildup of energy near the inner surface of the cloak.
While such an energy buildup exists for frequencies above the critical frequency (Figure \ref{fig:sim}B), it remains relatively localized in those cases.
At or below the critical frequency, though, the surface energy concentration extends across nearly the entire length of the inner boundary, meaning that an effectively uniform field surrounds the interior domain.
Since the elastic wave equations depend upon a Laplacian, they obey the mean value theorem, implying that such a uniform energy buildup is able to penetrate through any cloak, even a perfect one \cite{VC1,PC1,PC2,PC3}.

However, for a perfect cloak, the penetration would merely imply $\eta=0$, the presence of a negative efficiency implies that our SEC is still underperforming.
In particular, a perfect cloak would not produce the energy buildup that we observed in Figure \ref{fig:sim}.
As has been observed in other approximate cloaks \cite{EC1,EC4,AEC9}, imperfections can induce resonant modes within the cloak.
If we assume that the observed surface modes are resonantly excited by the SEC, we can predict a simple model for how the cloak should effect the response, using the standard 1D simple harmonic oscillator (SHO) dynamical response
\begin{equation}
u=\frac{F/m}{\sqrt{(\omega^2-\omega_c^2)^2+\omega^2\Gamma^2}}
\end{equation}
where $F/m$ is the effective acceleration and $\Gamma$ is the damping rate.
This would predict an efficiency of
\begin{equation}\label{eq:fit}
\eta^{SHO}(\Omega,Q)=1-\frac{1}{\sqrt{(\Omega^2-1)^2+\Omega^2/4Q^2}}
\end{equation}
where $\Omega=\omega/\omega_c$ is the dimensionless frequency and $Q=\omega_c/2\Gamma$ is the resonator quality factor.
This predicts an efficiency of 0 for $\Omega=0,2-1/4Q^2$, a minimum of $\eta=1-2Q$ for $\Omega=1$, and $\eta\to1$ for $\Omega\to\infty$, which is qualitatively similar to our calculated efficiencies for $\eta(E)$ (see solid curves in Figure \ref{fig:eff}).
For comparison, a perfect cloak efficiency should resemble a step function, jumping from 0 to 1 at $\Omega=1$.

While this harmonic oscillator model predicts the qualitative aspects of the SEC's performance, the exact value of $\omega_c$ is not captured by that model.
Since that determines the low frequency cutoff where the SEC's performance dramatically worsens, $\omega_c$ is critical to determining where the SEC is applicable and how it could be improved.
If we substitute the equations (\ref{eq:rhoTr}) and (\ref{eq:SEC}) into equation (\ref{eq:EOM}) and apply some simple mathematical operations (see supplement), we can derive an implicit series solution for $u$.
Notably, this implicit solution possesses a set of dimensionless parameters
\begin{equation}
\omega^2a^2(\frac{b}{b-a})^2\frac{1}{v_i^2}.
\end{equation}
Plugging in the parameters used in our simulations gives $\omega_{c,S}=$0.9286Hz and $\omega_{c,P}=$2.275Hz, which closely match our numerical results.
Notice too that
\begin{equation}\label{eq:OmC}
\omega_{c,i}=\frac{b-a}{b}v_i\frac{1}{a},
\end{equation}
which is equivalent to a wave with velocity $v_i(b-a)/b$ and wavelength $2\pi a$, i.e. a surface wave with wavelength equal to the circumference of the cloaked region.
Importantly, equation (\ref{eq:OmC}) implies that $\omega_c\propto1/a$, so increasing the size of the cloaked domain will lower the cutoff frequency.
To calculate $Q$ from first principles is a less trivial problem, so instead we treat it as a fitting parameter in our model and fit the efficiency characteristics of our SEC to get $\langle Q_S\rangle=1.57,\langle Q_P\rangle=3.89$.
In principle, for a real SEC, its performance could be measured by finding the resonant frequency and performance of the SEC at that frequency.
However, since no real material possesses the inhomogeneities of equations (\ref{eq:rhoTr}), (\ref{eq:SEC}), a real SEC would likely be constructed from constituent elements that approximate this (as in the standard cloak design \cite{RC1}).
As such, it should possess an upper cutoff frequency with wavelength of approximately the size of constituent element.
We therefore expect an SEC to be effective for blocking frequencies between $\omega_c\propto 1/a<\omega<\omega_{MM}\propto1/L$, where $\omega_{MM}$ is the characteristic frequency for a metamaterial of characteristic length $L$.
The design of seismic metamaterials is a relatively new field \cite{SMM1,SMM2,SMM3} but shows a great deal of progress, and when combined with techniques like additive manufacturing \cite{AM1}  it suggests the feasibility of realizing an SEC.
\textcolor{black}{Additionally, while the coordinate transformation used here necessitates the engineering of both stiffness and the density of the metamaterials, more complicated transformations exist that simplify the material requirements.
In particular, Ref. \cite{SpNlin} would result in a design that only required engineering the elasticity and would leave the density unchanged (i.e. uniformly equal to the surrounding medium's density).}

\begin{figure}\begin{center}
\includegraphics[scale=0.425]{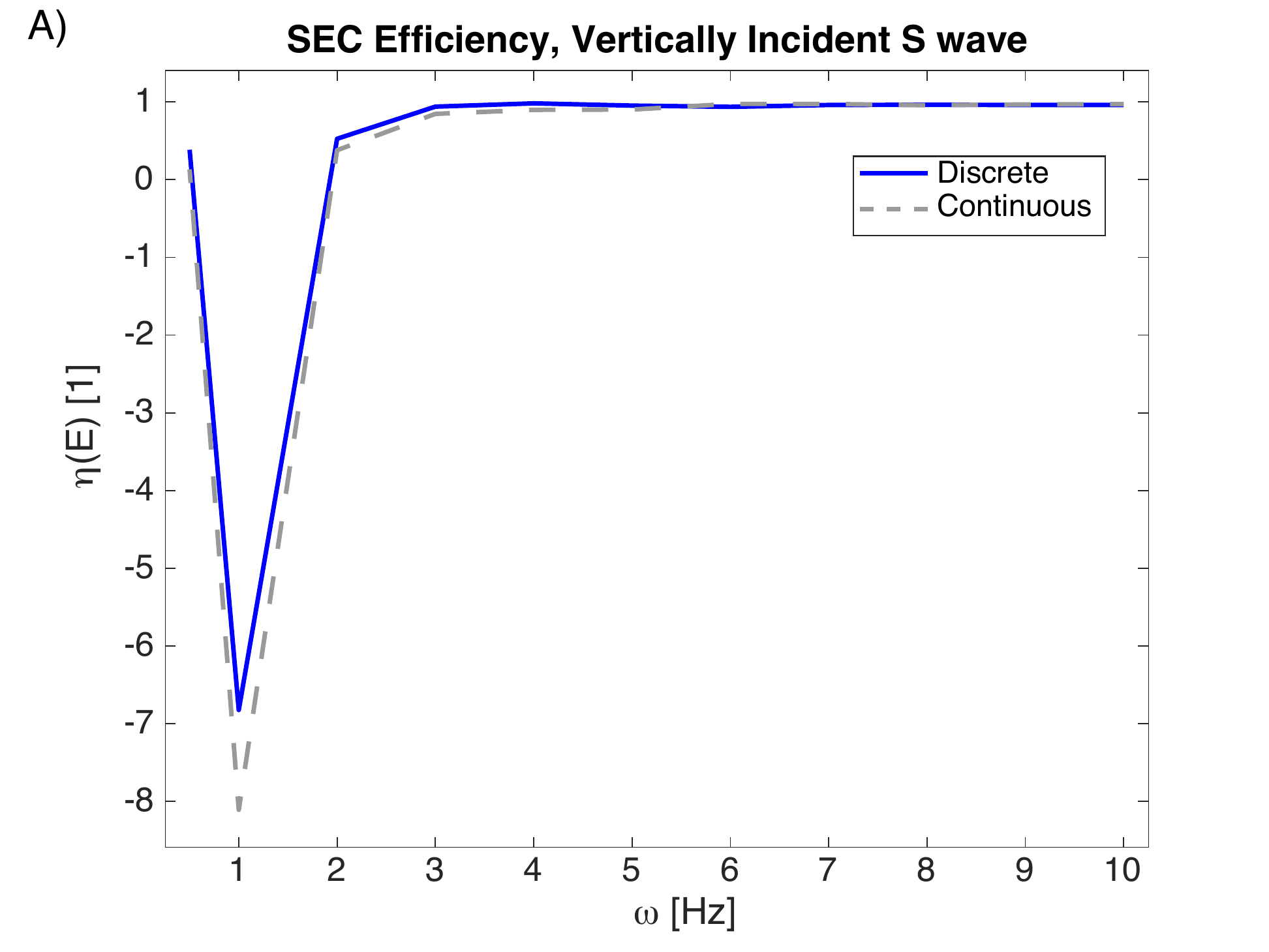}
\includegraphics[scale=0.425]{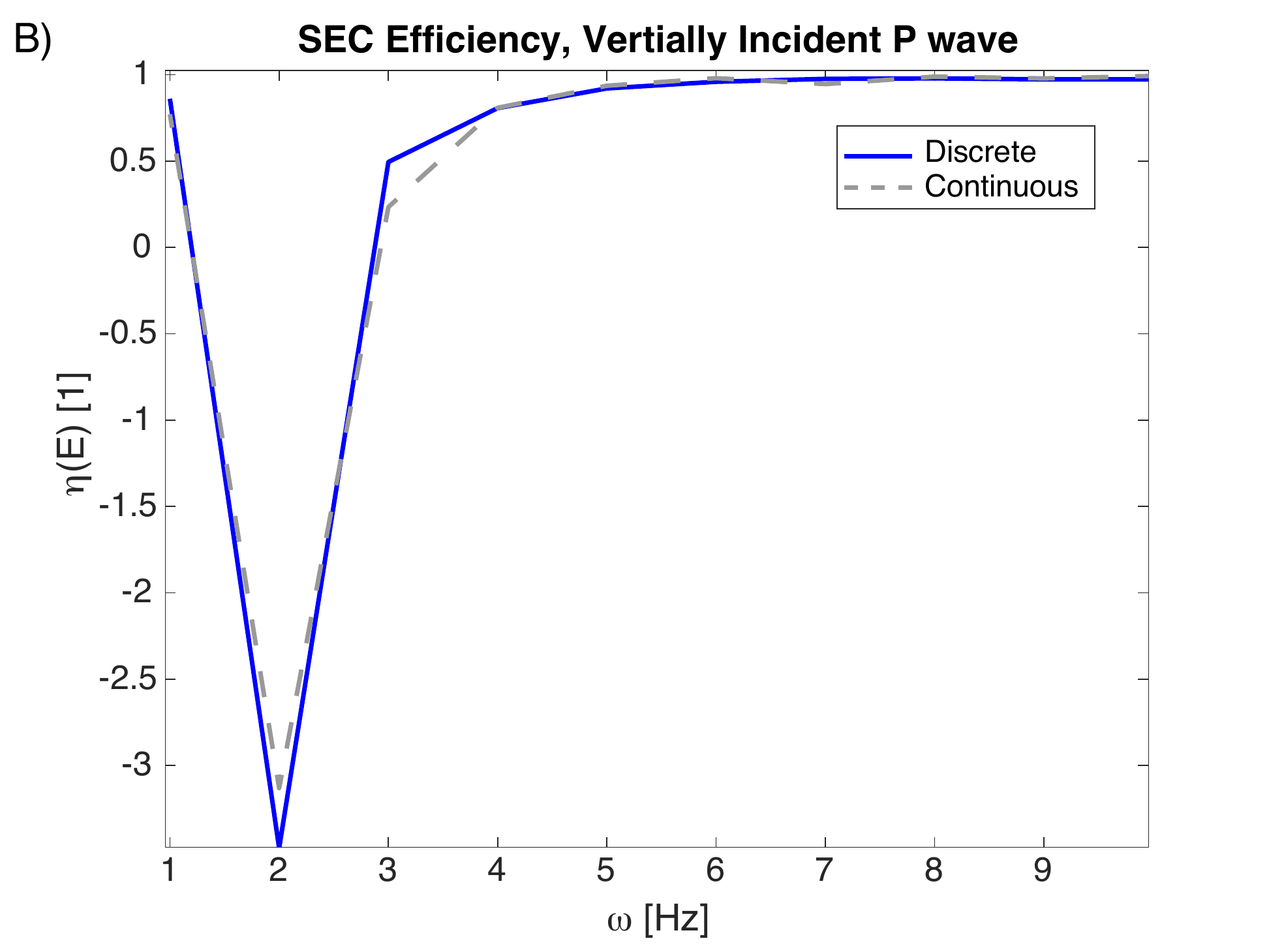}
\caption{\textcolor{black}{\label{fig:disc} Comparison of efficiency for the continuous (dotted grey line) and discrete (solid blue line) implementation of the SEC as a funcion of frequency. All plots are for vertical incidence. (A) Energy efficiency, S wave. (B) Energy efficiency, P wave.}}
\end{center}\end{figure}

\textcolor{black}{To determine the impact of this upper frequency cutoff and the feasibility of the SEC under more realistic conditions, we repeat our calculations using a discretized SEC.
As is the standard procedure in cloaking, we replace our inhomogeneous SEC given by equations (\ref{eq:rhoTr}) and (\ref{eq:SEC}) with a series of annular rings.
Each ring's density and elasticity tensor are selected to be the average value of these equations in their respective region.
In particular, we use 10 rings, each 1 meter thick, of uniform orthotropic material given by the mean value of the SEC in their 1 meter shells.
Although one meter is still quite thick, a thinner discretization scheme was avoided to ensure that the effects of discretization weren't disguised by an unrealistic number of layers and that the size of the rings was not significantly finer than the mesh grid used in our finite element simulations.
On the other hand, even a single layer would be too thin to display the predicted cutoff due to wave scattering on the discretized structure.
A 10 meter thick shell (i.e. a uniform layer of thickness equal to our SEC) would have an upper cutoff frequency of approximately 73 Hz, far above the 10 Hz practical limit that we employ (a shell engineered to be thin enough to be practical, then, would likely have an upper cutoff frequency in the $O(100)$ Hz range).
In this regime, then, we expect that the principle limitation to the discretized SEC's performance is not the geometric effect of the shell thickness but simply the deviation of the material properties from continuous limit.
As 10 layers typically shows good agreement with the continuous limit in experimental tests of metamaterial cloaks, we thus consider this case as the rigorous practical test of a realizable SEC.
Given the uniformity of the continuous SEC's performance under different angles of incidence, polarizations, and efficiency measures, we focus on the energy shielding of the discretized SEC for bulk S and P waves under vertical incidence.
We find in Fig. \ref{fig:disc} that the behavior is in good agreement with the continuous SEC case, even as the discretization removes the singular behavior of the cloak at the inner boundary.
In particular, we find clear agreement in the location of the lower cutoff frequency, qualitative wave dynamics, quantitative trend, and maximum efficiency.
Moreover, we do not observe any decrease in performance due to the discretization, even at high frequencies, as the frequency range of interest is far lower than the upper cutoff frequency.
We thus conclude that an SEC made from ordinary orthotropic material with tuned properties is an effective VIEE structure with much less limiting fabrication constriants.}

To summarize, we have developed a method of modulating the \textcolor{black}{challenge to realize} asymmetries of a perfect elastic wave cloak into a \textcolor{black}{more readily realized} symmetric elastic cloak.
By testing the performance of the cloak as a shield against seismic waves in a tunnel, we have demonstrated the feasibility of this design for VIEE, reducing the displacement within the cloaked tunnel by an order of magnitude for most of the frequencies corresponding to common seismic wave resonances (1-10Hz, \cite{HFW}) (compared to a tunnel with no cloak).
Since the frequency range where our approximate SEC fails is determined by the size of the cloaked region and the characteristic length of the SEC, switching to shield even larger regions ($a\gtrsim$100m) and using sufficiently small building blocks for the SEC ($L\lesssim$1m) should render this design quite effective for blocking seismic waves of any relevant frequency for a large structure.

Alternatively, a smaller SEC could be used specifically to focus upon protecting equipment and critical infrastructure against high frequency elastic waves (5-20+ Hz).
While these high frequency waves are often considered negligible in earthquake engineering due to their stronger dissipation, they can be dangerous in certain situations such as the survival of control equipment and circuits.
Geological conditions can lead to larger content of higher frequencies  \cite{Focus}.
Smaller structures can have their own higher resonant frequencies as well.
Since this can include internal resonances of walls, excitation of strong ground motion in soil, or the failure of critical infrastructure like water pipes, generators, etc. \cite{HFW,Soil}, the potential impact of high frequency waves can be immense and catastrophic.
Furthermore, as the failure modes of the SEC are focused around a critical frequency corresponding to surface elastic waves on the inner boundary of the SEC, the inclusion of damping or resonance shifting techniques practices in VIEE \cite{PVC} could be used to create a hybrid SEC/traditional earthquake abatement system with improved performance.
Moreover, as cloak designs have been developed for various geometries \cite{TO1,TO2,RC1,CC1} and could in principle be adapted to arbitrary geometric configurations, it is likely that this SEC approach could find ready application in a variety of critical structures.

\section{Acknowledgements}
We would like to thank Prof. Kathryn Matlack and Prof. Fatemeh Pourahmadian for their helpful discussion and suggestions.

\section{Supplementary Materials}
\subsection{Velocity and Acceleration Performance}
The SEC's efficiency for shielding against velocity or acceleration transmission closely matches its efficiency for shielding against displacement. This makes intuitive sense, as these quantities are expected to differ by factors of $\omega$, which cancel when we take the ratio of cloaked and uncloaked fields to calculate the efficiency.
\begin{figure}\begin{center}
\includegraphics[scale=0.35]{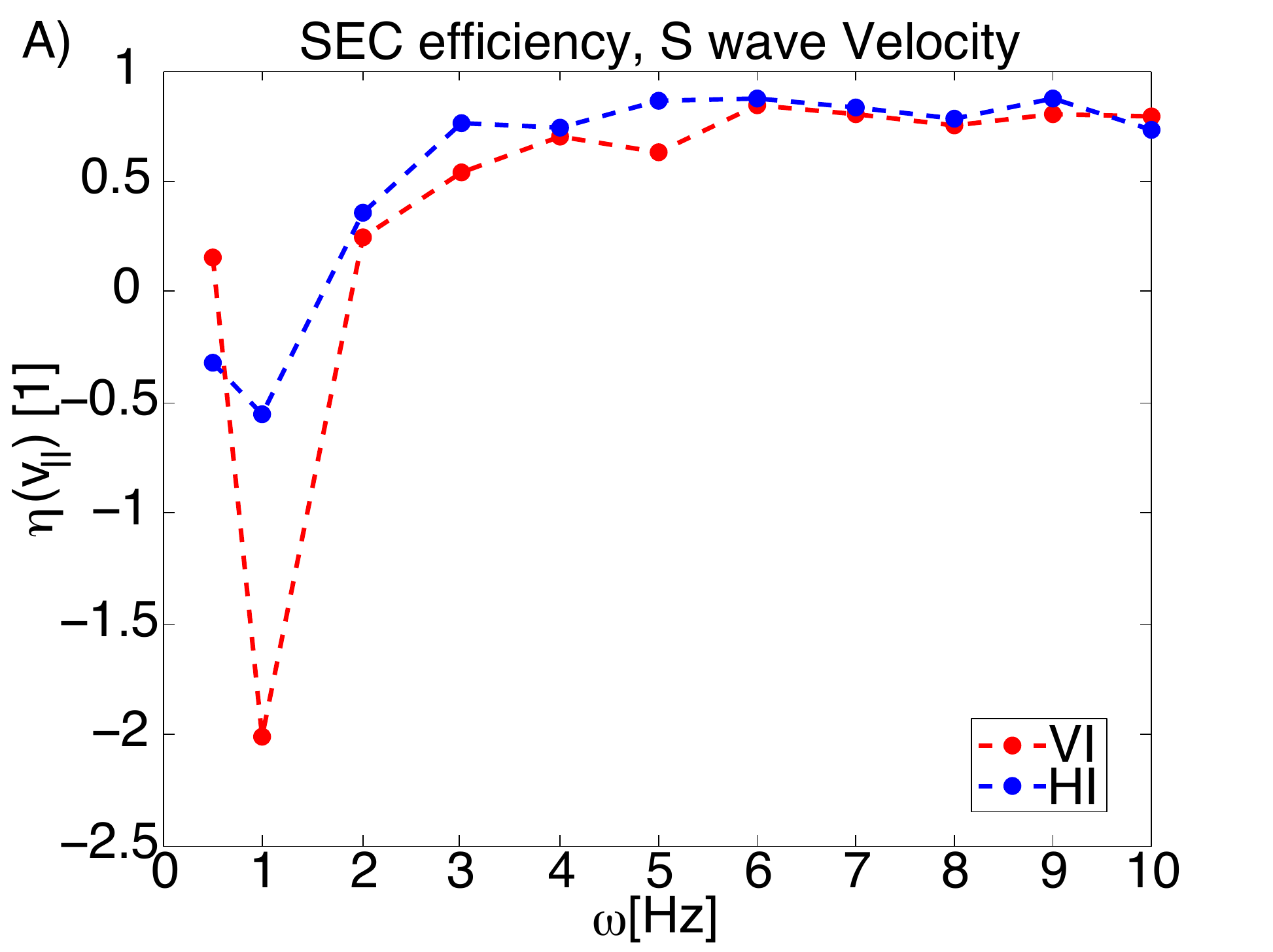}
\includegraphics[scale=0.35]{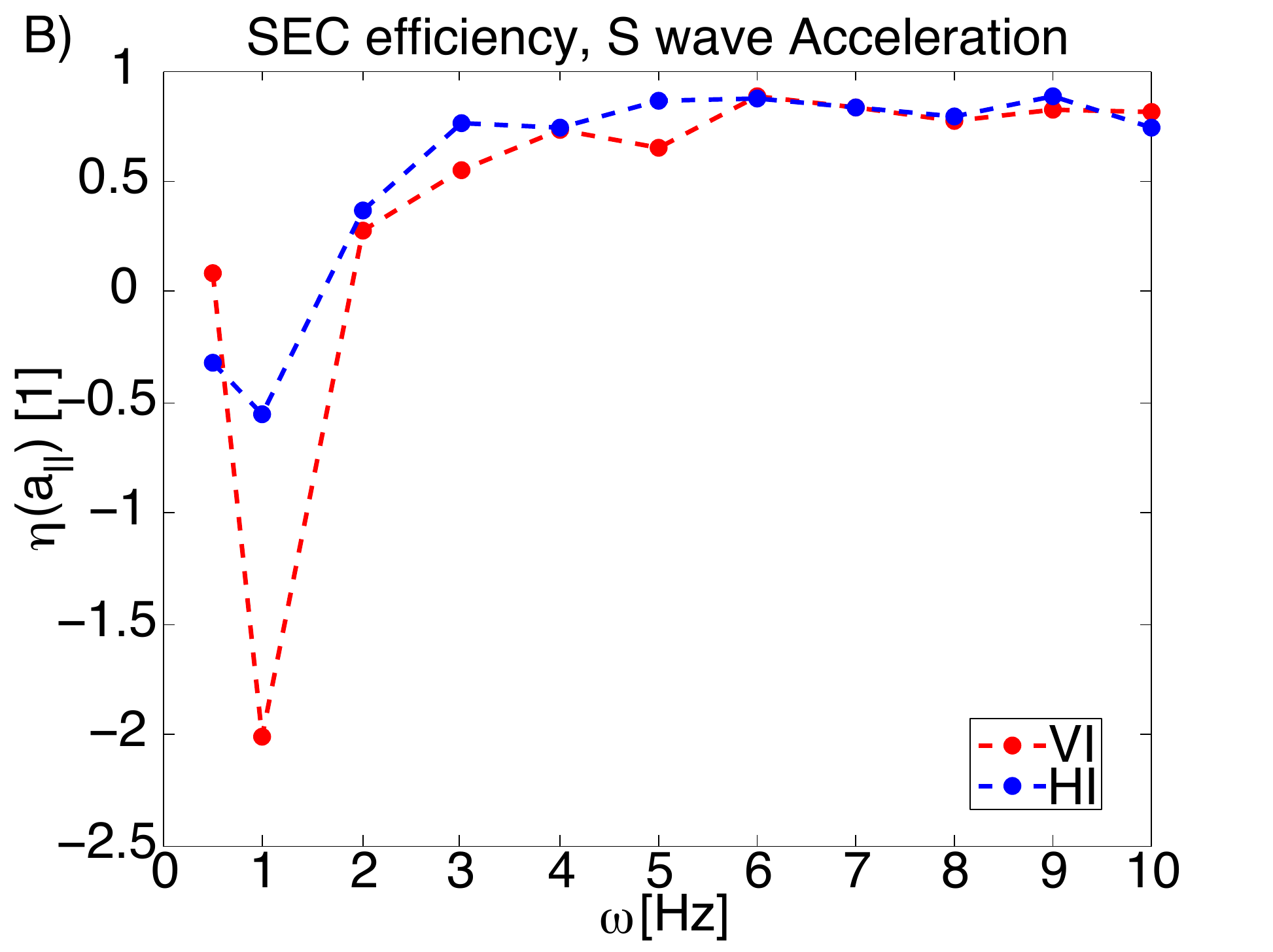}
\includegraphics[scale=0.35]{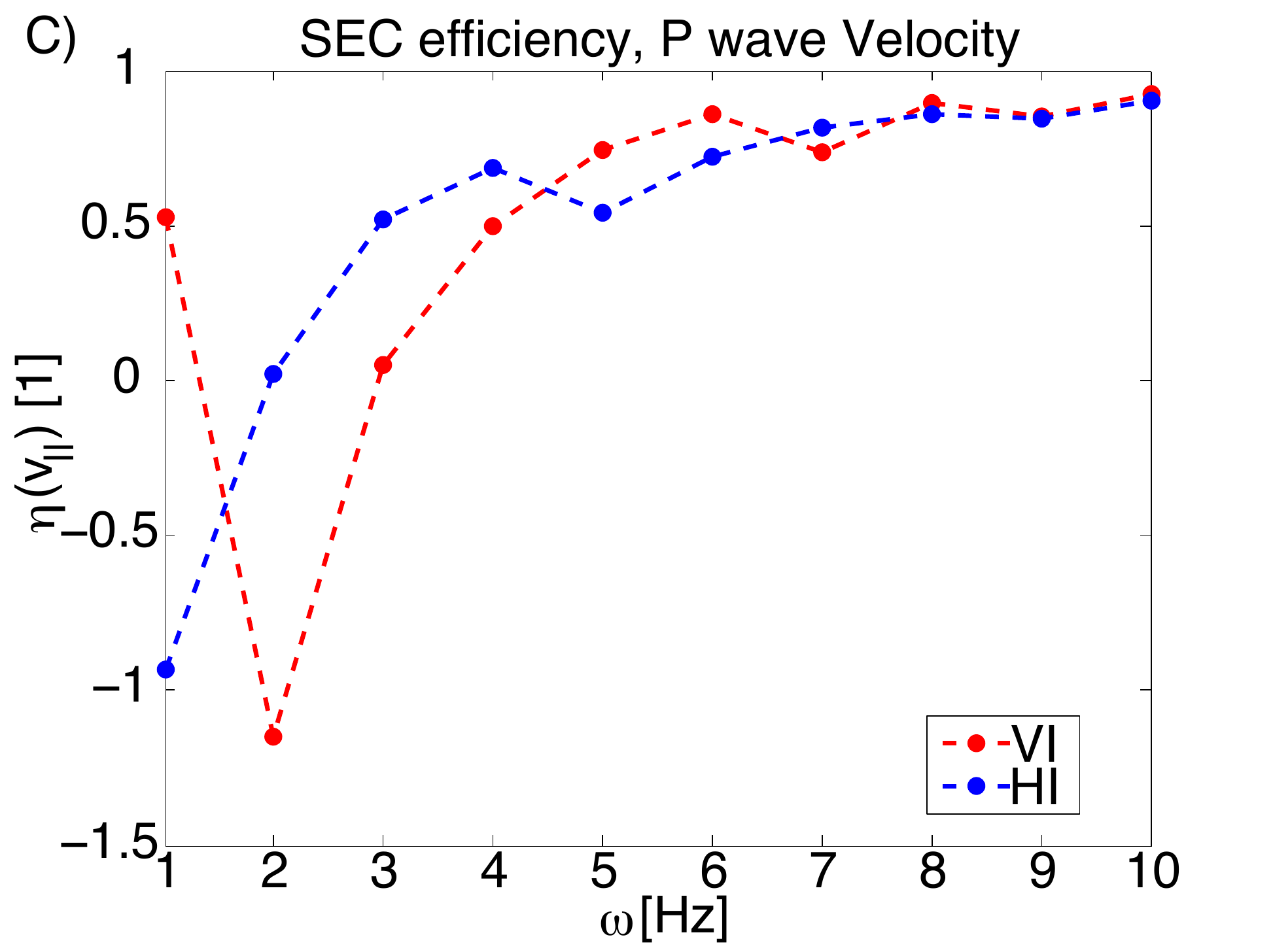}
\includegraphics[scale=0.35]{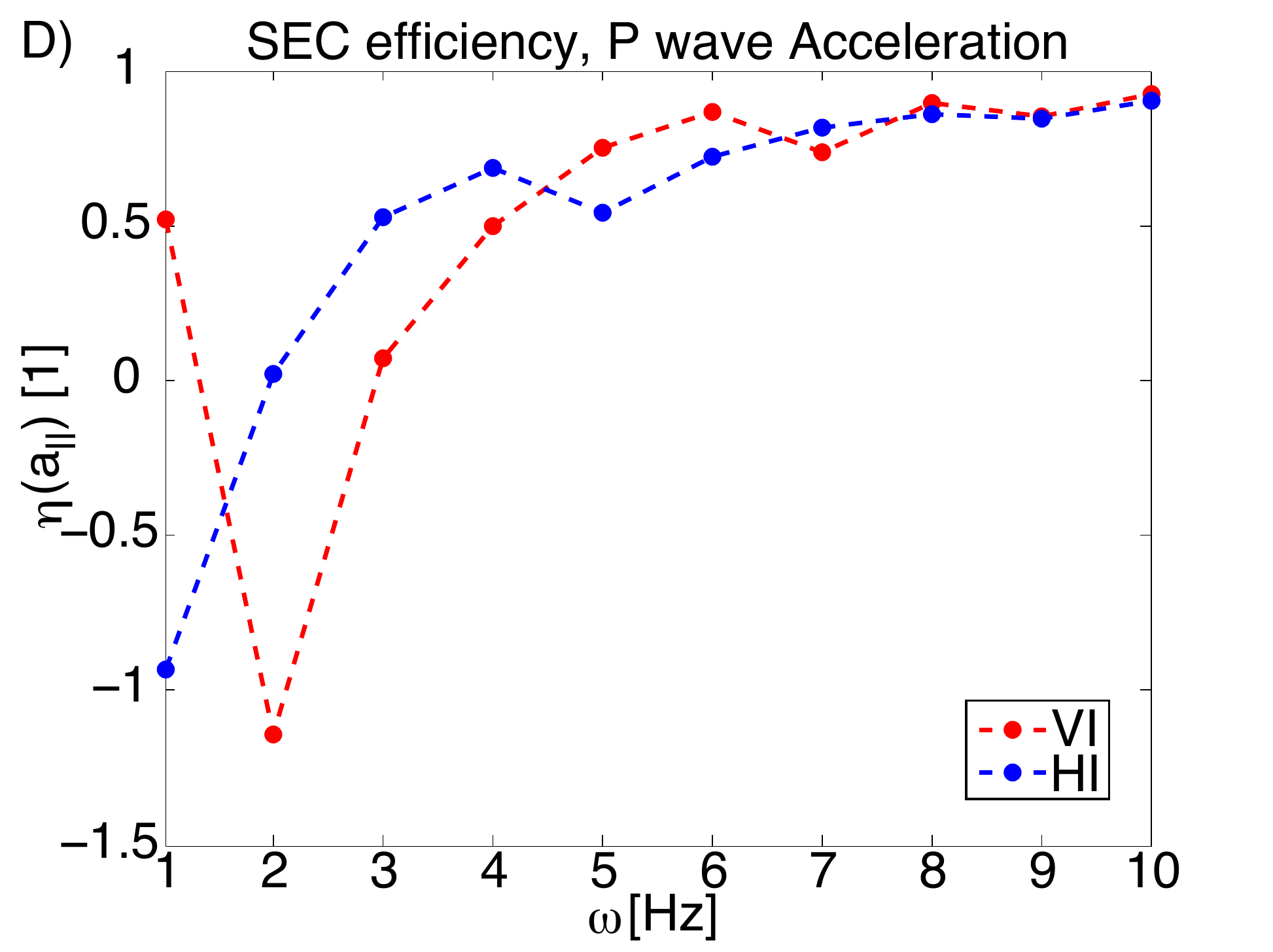}
\caption{Longitudinal component efficiency plots for the SEC as a function of frequency. Red dashed curve is data for vertical incidence (VI, wave from beneath the SEC), blue dashed curve is data for horizontal incidence (HI, wave from the left side of the SEC). (A) Velocity efficiency, S wave. (B) Acceleration efficiency, S wave. (C) Velocity efficiency, P wave. (D) Acceleration efficiency, P wave.}
\end{center}\end{figure}
\subsection{Transverse Component Efficiency}
While the overall trends for the efficiencies of the SEC for transverse components are similar to the longitudinal response of the main text, some deviations exist. However, these can be explained as arising from the relative magnitudes of the transverse and longitudinal components of the scattered waves, as only a tiny portion of the energy is relegated to these transverse modes, small fluctuations or numerical artifacts can have a large impact upon the calculated efficiency. Thus, these deviations from the longitudinal response are likely insignificant.
\begin{figure}\begin{center}
\includegraphics[scale=0.35]{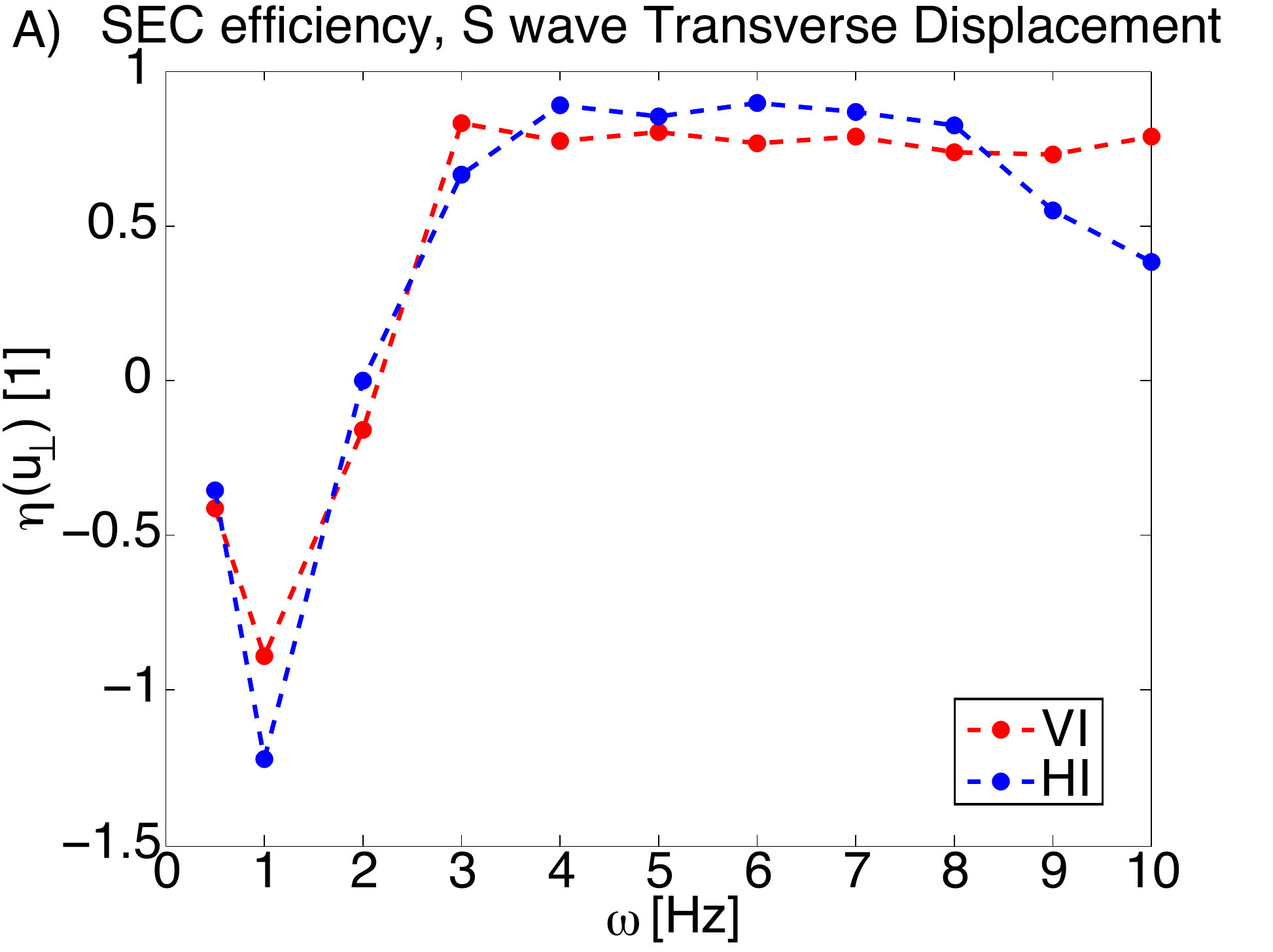}
\includegraphics[scale=0.35]{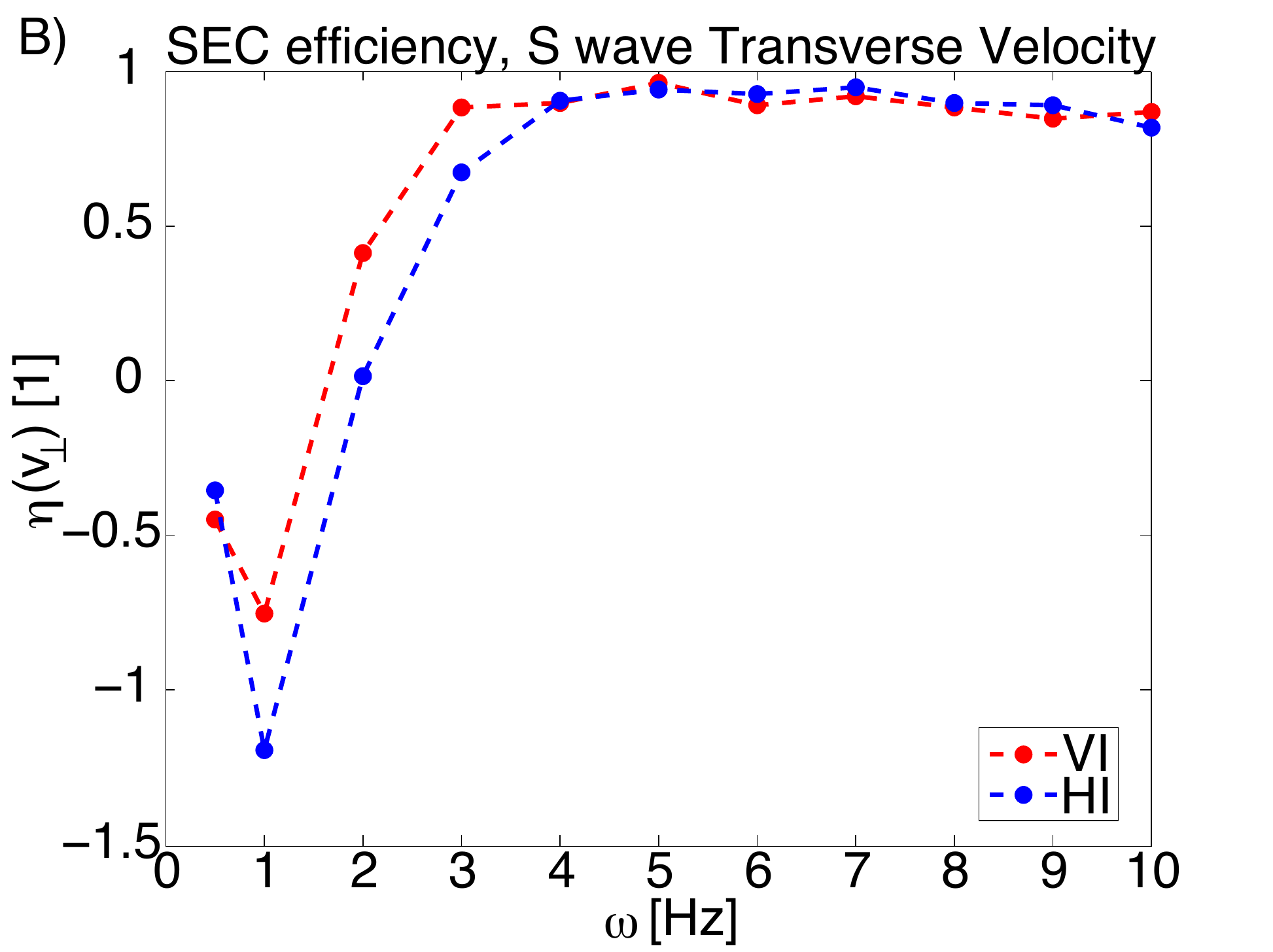}
\includegraphics[scale=0.35]{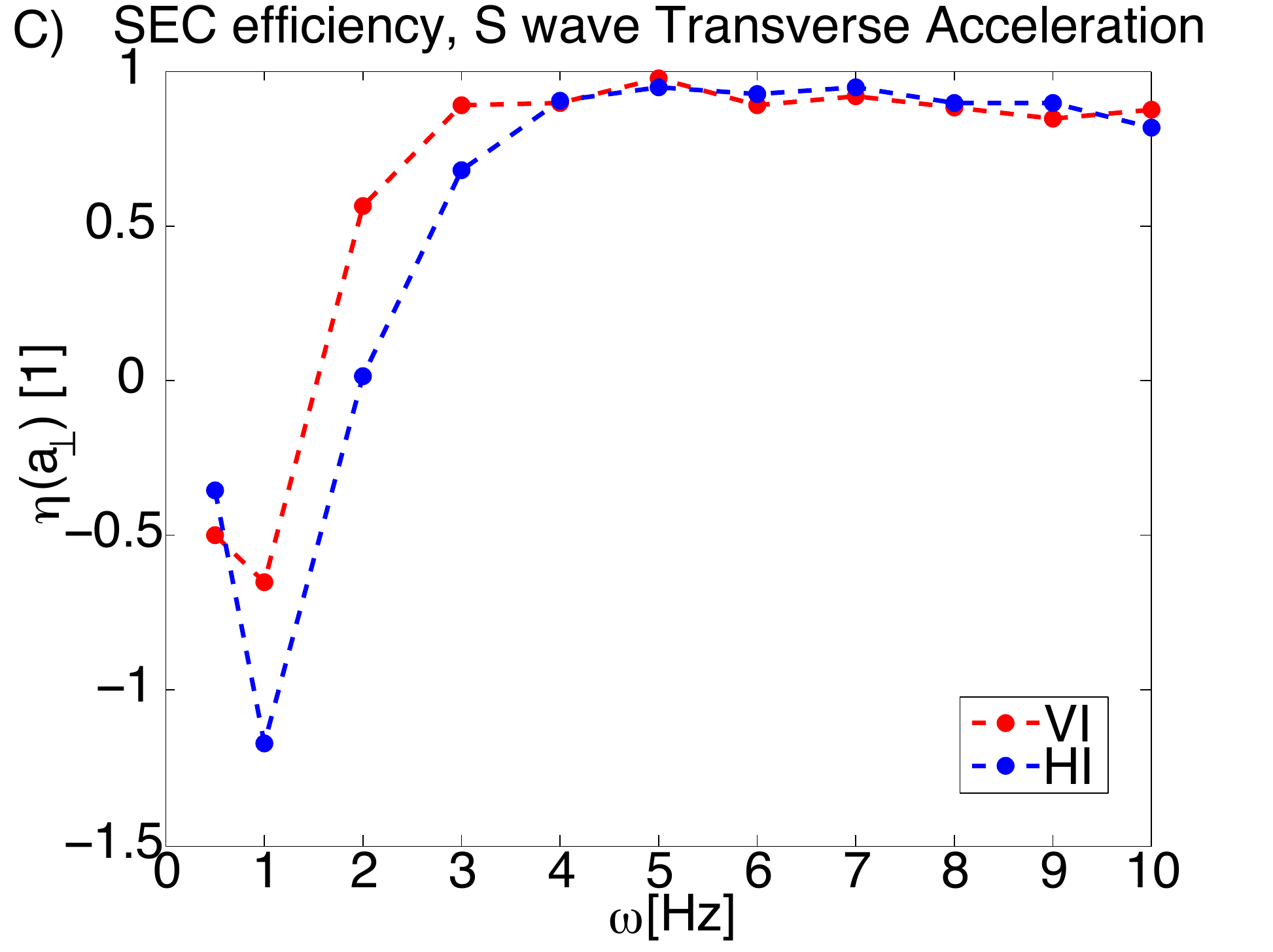}
\caption{Transverse component, S wave efficiency plots for the SEC as a function of frequency. Red dashed curve is data for vertical incidence (VI, wave from beneath the SEC), blue dashed curve is data for horizontal incidence (HI, wave from the left side of the SEC). (A) Displacement, (B) Velocity, (C) Acceleration.}
\end{center}\end{figure}
\begin{figure}\begin{center}
\includegraphics[scale=0.35]{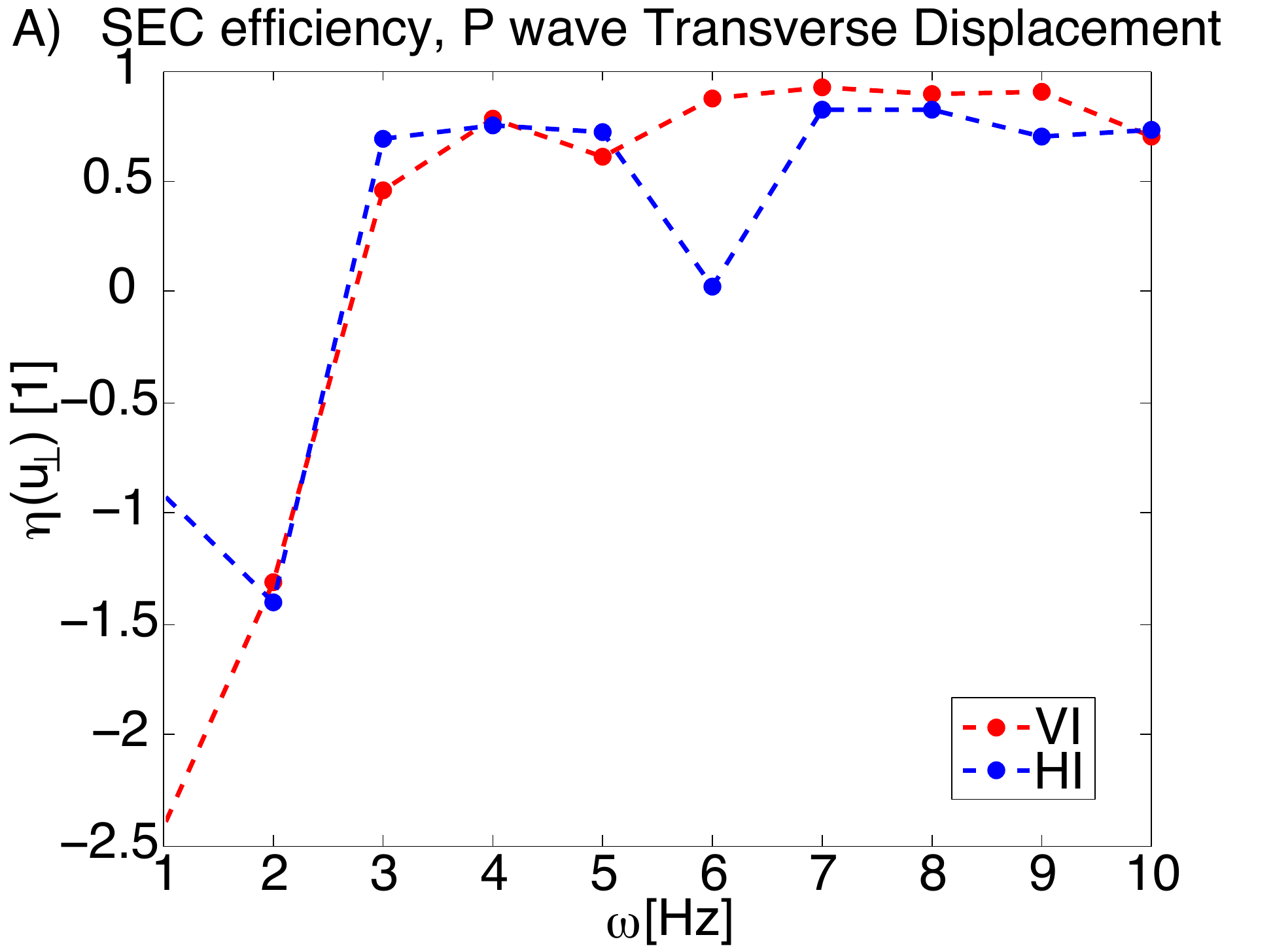}
\includegraphics[scale=0.35]{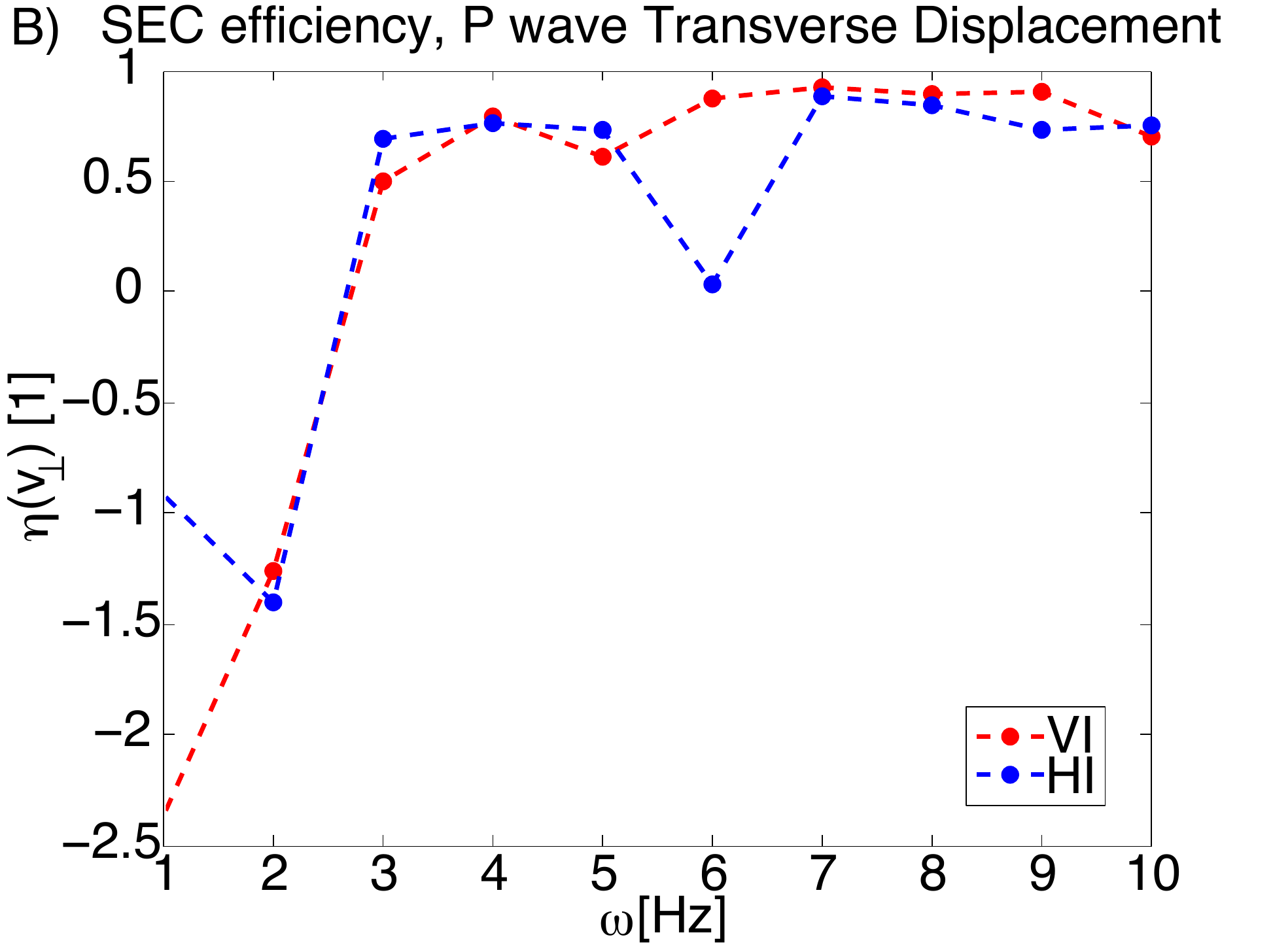}
\includegraphics[scale=0.35]{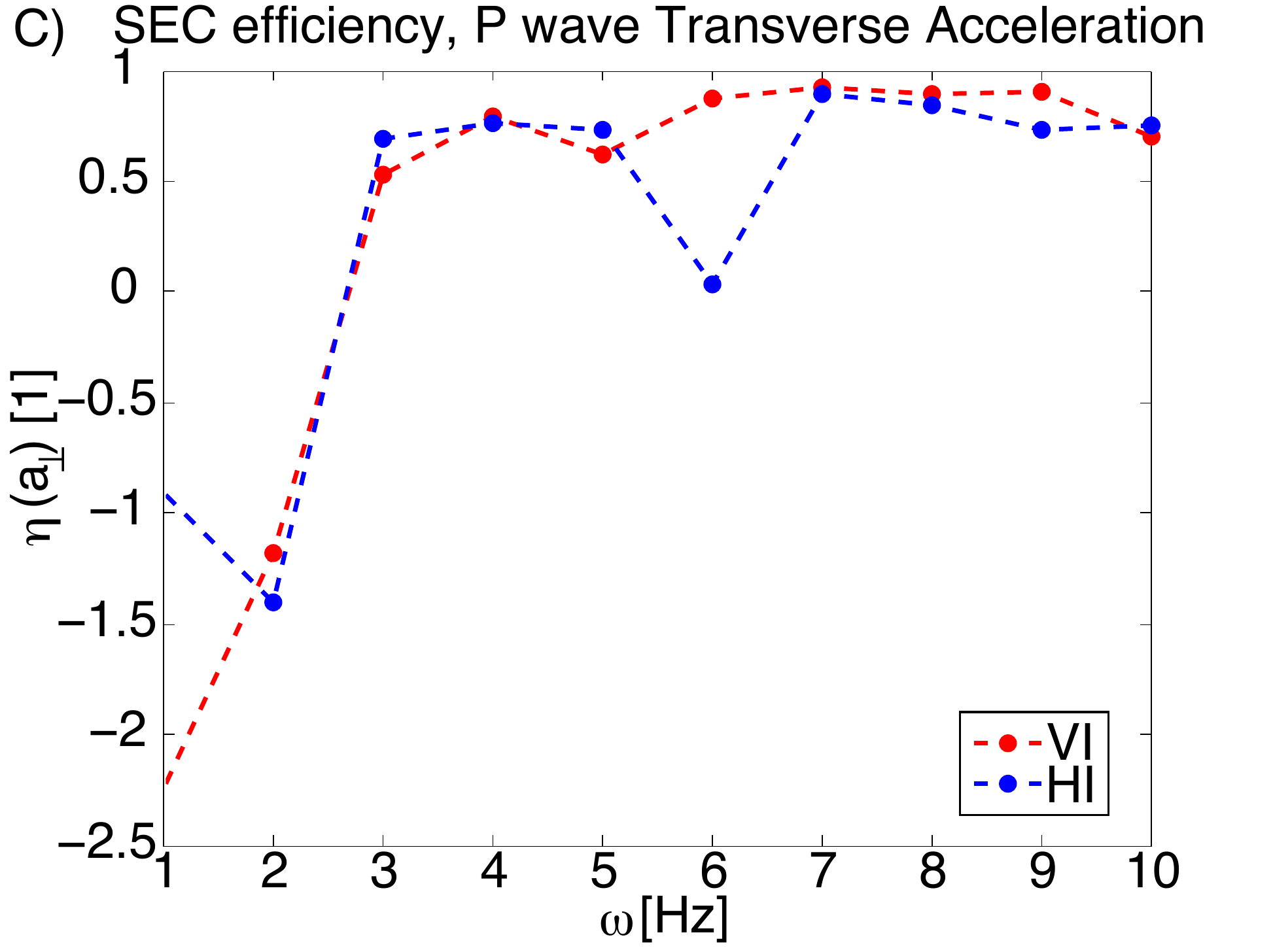}
\caption{Transverse component, P wave efficiency plots for the SEC as a function of frequency. Red dashed curve is data for vertical incidence (VI, wave from beneath the SEC), blue dashed curve is data for horizontal incidence (HI, wave from the left side of the SEC). (A) Displacement, (B) Velocity, (C) Acceleration.}
\end{center}\end{figure}
\subsection{Dimensionless Constants}
To find the dimensionless constants of the SEC we plug the parameters of equations (\ref{eq:rhoTr}) and (\ref{eq:SEC}) into equation (\ref{eq:EOM}).
We assume that variation along $z$ is negligible and thus specialize to polar coordinates, assuming a solution to the equations of motion
\begin{equation}
\vec{u}=\left(R_r(r)\hat{r}+R_\theta(r)\hat{\theta}\right)e^{i\omega t+in\theta}
\end{equation}
where $\omega,n$ are the temporal and azimuthal eigenvalues under separation of variables.
Plugging these in and multiplying the equations of motion by $r(r-a)$ gives
\begin{gather}
-\omega^{2}(r-a)^{2}(\frac{b}{b-a})^{2}\frac{\rho}{\lambda+2\mu}R_{r} =(r-a)^{2}R_{r}^{\prime\prime} \nonumber\\ +(r-a)R_{r}^{\prime}-R_{r}-(1-\frac{a}{r})n^{2}\frac{\mu}{\lambda+2\mu}R_{r} \\
+in\left(\frac{\lambda+\mu}{\lambda+2\mu}(r-a)R_{\theta}^{\prime}-R_{\theta}-\frac{\mu}{\lambda+2\mu}(1-\frac{a}{r})R_{\theta}\right) \nonumber
\end{gather}
and
\begin{gather}
-\omega^{2}(r-a)^{2}(\frac{b}{b-a})^{2}\frac{\rho}{\mu}R_{\theta} = r(r-a)R_{\theta}^{\prime\prime}\nonumber\\
+(r-a)R_{\theta}^{\prime}-(1-\frac{a}{r})R_{\theta}-n^{2}\frac{\lambda+2\mu}{\mu}R_{\theta}\\
+in\left(\frac{\lambda+\mu}{\mu}(r-a)R_{r}^{\prime}+\frac{\lambda+2\mu}{\mu}R_{r}+(1-\frac{a}{r})R_{r}\right)\nonumber
\end{gather}
which can be solved by the method of Frobenius for $R_r,R_\theta$.
Notably, though, the only terms which are independent of $r$ or $d/dr$ possess dimensionless parameters which can be used to define the scale of different variables.
Some of these (e.g. $n^2\mu/(\lambda+2\mu)$) express obvious relations (e.g. the ratio of the speeds of sound times $n^2$), but
\begin{equation}
\xi_i=\omega^{2}a^{2}\left(\frac{b}{b-a}\right)^{2}\frac{1}{v_{i}^{2}}
\end{equation}
relates the frequency of the incoming waves to the natural length scale of the cloak and is thus a non-trivial parameter for this problem.

\end{document}